\documentclass[fleqn,usenatbib]{mnras}
\usepackage [caption=false]{subfig}
\let\oldAA\AA
\renewcommand{\AA}{\text{\normalfont\oldAA}}

\usepackage{newtxtext,newtxmath}

\usepackage[T1]{fontenc}

\DeclareRobustCommand{\VAN}[3]{#2}
\let\VANthebibliography\thebibliography
\def\thebibliography{\DeclareRobustCommand{\VAN}[3]{##3}\VANthebibliography}

\usepackage{graphicx}	
\usepackage{amsmath}	

\title[Disk-jet coupling in AGN]{Disk-jet coupling across the spectral transition in supermassive black holes}

\author[Kang et al.]{Jia-Lai Kang$^{1,2,3}$\thanks{jialai.kang@durham.ac.uk},
Chris Done$^{1,4}$\thanks{chris.done@durham.ac.uk},
Scott Hagen$^{1}$,
Mai Liao$^{5}$,
Matthew J. Temple$^{1}$,
\newauthor
John D. Silverman$^{4,6,7,8}$,
Junyao Li$^{9}$,
Jun-Xian Wang$^{2,3}$
\\
$^{1}$Centre for Extragalactic Astronomy, Department of Physics, Durham University, South Road, Durham, DH1 3LE, UK\\
$^{2}$Department of Astronomy, University of Science and Technology of China, Hefei 230026, China \\
$^{3}$School of Astronomy and Space Science, University of Science and Technology of China, Hefei 230026, China \\
$^{4}$Kavli Institute for the Physics and Mathematics of the Universe (Kavli IPMU, WPI), UTIAS, Tokyo Institutes for Advanced Study, University of Tokyo, \\ Chiba, 277-8583, Japan\\
$^{5}$Universidad Diego Portales, Av Republica 180, Santiago, Reg\'i\'on  Metropolitana, Chile\\
$^{6}$Department of Astronomy, School of Science, The University of Tokyo, 7-3-1 Hongo, Bunkyo, Tokyo 113-0033, Japan \\
$^{7}$Center for Data-Driven Discovery, Kavli IPMU (WPI), UTIAS, The University of Tokyo, Kashiwa, Chiba, 277-8583, Japan \\
$^{8}$Center for Astrophysical Science, Department of Physics \& Astronomy, John Hopkins University, Baltimore, MD 21218, USA \\
$^{9}$Department of Astronomy, Univerity of Illinois at Urbana-Champaign, Urbana, IL 61801, USA \\
}
\date{Accepted XXX. Received YYY; in original form ZZZ}

\pubyear{\the\year{}}

\begin{document}
\label{firstpage}
\pagerange{\pageref{firstpage}--\pageref{lastpage}}
\maketitle

\begin{abstract}

{
Accretion flows in both stellar and supermassive black holes show a
distinct spectral transition. This is seen directly in binaries and
changing look AGN, and also in a recent sample of eROSITA X-ray
selected, unobscured AGN where the stacked spectral energy
distributions (SEDs) for a single black hole mass bin 
($\log M/M_{\sun} =8-8.5$) clearly show the UV bright disk 
appearing as the luminosity increases. In binaries, this 
transition is associated with
a change in radio jet, from coupling to the X-ray hot flow with
$L_{\rm R}\propto L_{\rm X}^{0.7}$ (Fundamental Plane relation), to collapsing
when the X-ray hot flow collapses into a disc.  We explore the radio
behaviour across the transition in our AGN sample by stacking VLASS
images.  We significantly detect weak radio emission even after
subtracting the contribution from star formation in the host
galaxy. The residual radio emission remains relatively constant across
the transition, despite the mean mass accretion rate changing by a
factor 6 and UV flux changing by a factor 100. However, the X-rays
change by only a factor 2, giving a constant radio to X-ray flux ratio
as predicted by the `fundamental plane'. We show that this is
consistent with these AGN having the same compact radio jet coupling
to the X-ray hot flow (not the disc) as in the binaries. The most
significant difference is the persistance of the coronal X-rays across
the spectral transition in AGN, whereas in binaries the coronal X-rays
can be very weak in the disc dominated state.
}
\end{abstract}

\begin{keywords}
accretion, accretion discs -- black hole physics -- galaxies: active
\end{keywords}

\section{Introduction}

Accretion flows onto stellar mass black hole binary systems (BHXRBs) show a remarkable spectral transition, from being dominated by an optically thick thermal component peaking at 1--2~keV at high luminosity (soft state), to optically thin, Comptonised emission peaking at 100~keV at lower luminosity (hard state). 
This is most easily interpreted as a change in the nature and geometry of the accretion flow, from a standard disc to a radiatively inefficient/advection dominated accretion flow (ADAF/RIAF), especially as this transition occurs at $\sim 0.01L_{\rm Edd}$ for a slow change in mass accretion rate \citep{Maccarone03,Vahdat_Motlagh19}, which is the maximum luminosity for an ADAF \citep{Narayan94}. 
Unlike the thermal disc, whose temperature changes with black hole mass 
at a given Eddington fraction ($L/L_{\rm Edd}$), the ADAF properties are mostly scale invariant. This predicts the transition should be present at similar luminosity in the accretion flows of the supermassive black holes which power AGN and Quasars \citep{Narayan94}. There is clear evidence for this in the ``Changing Look'' AGN, where a single AGN fades or brightens, giving a change in the broad band ionizing spectral energy distribution and consequent broad line emission as it crosses this luminosity \citep{Noda18,Ruan19,Zeltyn24, Jana_2025}.
Nonetheless, these are rare systems \citep[e.g.,][]{Temple_2023}. More compelling evidence would be to see this transition in the broader population. However, this is difficult as an AGN dims towards $L\sim 0.01L_{\rm Edd}$ as the host galaxy emission becomes more important, and the visibility of a UV disc and broad line region can be easily suppressed through obscuration. Recently, these issues were tackled in \citet{Hagen24} by building a new X-ray selected sample from eROSITA \citep{Brunner_2022, Liu_2022}, with $N_{\rm H}<10^{22}$~cm$^{-2}$ such that reddening associated with cold gas is small. This was crossmatched with galaxies imaged in the HyperSuprime Cam field \citep{Aihara_2022, Li_2024}, so that the extended host galaxy emission could be modelled and subtracted. The host stellar mass also gave an estimate for black hole mass across the entire sample, removing the requirement that the broad emission lines are detected. Selecting only a single mass bin ($\log M_{\rm BH}/M_{\sun} =8.0-8.5$)
clearly shows that the emission from a blue optically thick disk--like flow collapses at $L\sim 0.01L_{\rm Edd}$ in the entire population \citep{Hagen24} and that this has even more impact on the UV emission and its reprocessing into broad emission lines (BLR) \citep{Kang_2025}.

In binaries, this spectral transition is associated with a dramatic change in the radio emission. The radio flux from the compact steady jet correlates with the X-ray emission from the hot flow in the hard states, with $L_{\rm R}\propto L_{\rm X}^{0.7}$ \citep[see e.g.][]{Corbel_2013}. This correlation breaks down at the 
hard-soft transition, where discrete ballistic jet ejections dominate the radio \citep[see e.g.][]{Gallo_2003}. The correlation also breaks down after the transition, where the X-rays in BHXRBs are dominated by the disc component \citep{Gallo_2003, Fender_2004}. 
This shows that the radio emission correlates with the X-ray hot coronal flux, and not with the total mass accretion rate which powers the disc component. 
This is consistent with models where the X-ray hot flow not the thin disc connects to the jet, plausibly because the flow has large scale height so supports large scale height magnetic fields which provide the poloidal component close to the black hole which powers the jet. The radio can then be 
produced by the sum of self-absorbed synchrotron emission components from a vertically extended \citet{Blandford_1979} conical jet \citep{Merloni_2003,Heinz_2003}. 
This predicts a `fundamental plane of radio-X-ray emission' (hereafter FP) for the hard state emission where $\log L_{\rm R}=0.7\log L_{\rm X} +0.7\log M +C$, connecting across from BHXRBs to AGN (see \S\ref{sec:FP}). The mass dependence is important as the BHXRBs 
all have similar black hole masses (within a factor 2--3), which reveals the correlation directly, but AGN span from $10^5-10^{10}M_\odot$, which leads to considerable scatter in a direct correlation \citep{Merloni_2003}.

Multiple studies have tried to associate the very wide range of radio behaviour seen in AGN with this complex behaviour seen in the BHXRBs states
\citep{Kording_2006, Sobolewska_2011, Svoboda_2017, Fernandez_2021, Moravec_2022}. A small fraction ($\sim 10\%$) of AGN have powerful relativistic jets, but unlike BHXRBs jets (either compact or ballistic), these have high bulk Lorentz factor of $\sim 10-20$ \citep{Ghisellini_2010}. This means that the jet emission is strongly beamed and dominates the radio and X-ray bandpasses where the jet aligns with the line of sight (Blazars: BL Lacs and Flat spectrum radio sources: FSRQ). The relativistic jet is still clearly evident when sources are 
misaligned, as it powers strong radio emitting extended structures (Fanaroff-Riley types 1 and 2, \citealt{urry_1995}). However, most AGN do not have such large scale/high bulk Lorentz factor/high power jets, but do have compact emission which could be from a steady jet such as is seen in the 
hard state BHXRBs \citep{Falcke_1995, Giroletti_2009, Panessa_2013}. However, this origin is controversial as there could also be contributions from shocks from AGN winds \citep{Zakamska_2014,Nims_2015,Chen_2024}, or radio emission from the X-ray coronae itself \citep{Laor_2008, Behar_2015,Chen_2023}, and/or star formation processes \citep{Condon_1992, Thean_2001, Delvecchio_2017}. See e.g. \cite{Panessa_2019} for a review. 

The distinction between sources with and without a high Lorentz factor 
relativistic jet seems more robust than the radio-loud (RL) / radio-quiet (RQ) distinction. Radio-loudness $R$, is defined as the
ratio between the radio and optical fluxes \citep[e.g.,][]{Kellermann_1989}, with $R=10$ used as the boundary. AGN with a powerful relativistic jet are indeed very RL, but the drop in optical luminosity at the spectral transition could also lead to a compact jet with much smaller radio luminosity/smaller bulk Lorentz factor like that of the BHXRBs being classed as RL \citep{Kording_2006, Sobolewska_2011, Svoboda_2017}. 

Here we take the single mass range AGN sample of \citet{Hagen24,Kang_2025} and extend the SEDs into the radio band to investigate the effect of the spectral transition on the properties of the jet.
Only 29/1305 of our sources are detected individually (these are RL) 
so we stack the undetected sources into three bins 
of increasing $\dot{m}=L/L_{\rm Edd}$ (defined by the optical/UV/X-ray SEDs) to explore their faint radio emission. We subtract the radio emission from a matched host galaxy sample, and find that the residual radio emission is relatively constant, despite spanning the spectral transition. The 
mass accretion rate changes by a factor 6 from the first bin below the spectral transition to the brightest bin, and there is a factor 100 change in UV flux. However, the coronal X-ray luminosity changes by less than a factor 2 across the SEDs, unlike the BHXRBs where the coronal X-ray luminosity can drop to very low values after the transition. 
The lack of change in radio emission is then consistent with the (lack of) change in X-ray emission, and the resulting radio-X-ray ratio is broadly consistent with the FP relation of \citet{Merloni_2003}, but the large change in optical/UV means that the stacked spectra switch from RL to RQ across the spectral transition. This shows that the FP i.e. radio to X-ray ratio, gives more insight into the jet properties in AGN than the classical radio to optical ratio, again showing that, similar to the BHXRBs, the radio follows the X-ray corona rather than the total mass accretion rate. Thus the term `disk-jet' coupling which is often used in the literature is a somewhat of a misnomer, as the jet couples with the X-ray hot flow, not to the optically thick disc. 

This supports FP models where the 
origin of the radio emission in these RQ AGN is from a low power, steady compact jet with low/moderate bulk Lorentz factor, which couples to the X-ray emitting accretion flow in a similar way to that seen in the stellar mass black hole binaries. 

There are some differences though: in 
the stellar mass BHXRBs the X-ray coronal emission can be very low after the transition, while in AGN the X-ray flux stays at roughly the level of the brightest ADAF \citep{Kubota_2018}. 
Another difference is that the AGN clearly show very significant dispersion with respect to the FP, with a tail to higher core radio to X-ray ratios. We suggest that distance from the FP is a better way to characterise strong jet power sources, and speculate that this is due to some AGN reaching higher black hole spins than BHXRBs due to their very different formation and evolution histories.

\section{eFEDS-HSC sample}

\subsection{Stacked VLASS images}

\par The eFEDS field is fully covered by the Karl G. Jansky Very Large Array Sky Survey \citep[VLASS,][]{Lacy_2020}. This radio survey is conducted with the NRAO Very Large Array (VLA) in its B-configuration within 2--4 GHz band. VLASS images have a pixel size of 1.0\arcsec, spatial resolution of 2.5\arcsec, and a typical rms of 120 $\mu$Jy beam$^{-1}$ for a single-epoch image. All the three epoch Quick Look images of the eFEDS field have now been released, giving a combined typical rms of $\sim$ 70 $\mu$Jy beam$^{-1}$. For each source, we download the three epoch cut-out images from the Canadian Astronomy Data Centre (CASC) database, with a size of 81 $\times$ 81 arcsec$^{2}$.

For each source, we median stack the three VLASS images and perform the source detection on the stacked image. We adopt a simple criterion for detection, i.e., the peak flux within the center $3\arcsec \times 3\arcsec$ area is larger than 420 $\mu$Jy (the typical rms of a stacked image is 69 $\mu$Jy, thus $\sim$ 6$\sigma$ level). By combining the three epoch images, we detect 29 sources, 16 of which are in the VLASS Quick Look epoch 1 catalog. Most of them are point-like sources as shown in Figure \ref{fig:indi_SED}, and most are radio-loud with radio loudness $R > 10$, where $R = \frac{f_{5 \rm~GHz}}{f_{2500 \AA}}$ \citep[e.g.,][]{Kellermann_1989, Schulze_2017}. These detected sources are fairly evenly distributed across the sample in terms of $\dot{m}$ (see Figure \ref{fig:indi_SED}), with detection rates of 2.9\%, 2.0\% and 2.3\%, respectively. This is substantially below the often quoted $\sim 10$\% of AGN being RL. 
The VLASS flux limits predicted from the SEDs are 243, 39 and 6 $\mu$Jy for $R=10$ at the mean redshift, for bright, middle and faint bins respectively, so we could be missing some RL sources, especially in the fainter bins. 

\par We visually inspect all the remaining sources, and find 55 with contamination from nearby sources or spurious features (see Figure \ref{fig:vlass_offset}). These 84 sources (detected and contaminated) are excluded from the remaining analysis.

\par We then stack the remaining undetected source cut-out images following the procedure in \citet{Liao_2022, Liao_2024}. In brief, we use the coordinates of the optical counterparts in the HSC catalog to align the cutouts and median stack them on a pixel-by-pixel basis to obtain a stacked image \citep[e.g.,][]{White_2007, Fawcett_2020}. The typical positional accuracy of the VLASS quick look images is $\lesssim 0.5\arcsec$\footnote{\url{https://safe.nrao.edu/wiki/pub/JVLA/VLASS/VLASS_Memo_013_Epoch_1_Quicklook_Data_Release.pdf}}, while the astrometry of HSC is much better with typical error $< 0.05\arcsec$. The uncertainties of both positions are much smaller than the spatial resolution of VLASS images ($\sim$ 2.5\arcsec), and should not induce significant biases on the stacked fluxes.
We adopt median stacking which is less sensitive to outliers (e.g., a small fraction of more radio-loud objects despite our removal of all detected sources) compared with mean stacking.
Since the varying point-spread-functions (PSFs) of the VLASS images have not been \textsc{clean}ed, we only adopt the peak flux density from the central pixel in each stacked image, where the nominal point-like sources should be centered \citep{White_2007}, as an average radio flux for analysis, and we use the bootstrapping method in \citet{Karim_2011} to derive uncertainties.

\par To achieve high enough signal-to-noise ratios, we rebin the original eight luminosity bins in \citet{Hagen24, Kang_2025} into three bins, faint ($\nu L_{3500}$ in [42.1, 43.3], 469 sources, top row), middle ($\nu L_{3500}$ in [43.3, 44.1], 580 sources, 2 out of 3 middle row) and bright ($\nu L_{3500}$ in [44.1, 45.3], 256 sources, bottom row and the brightest middle row bin). In Figure \ref{fig:sed} we show the re-binned SED of the three bins. Instead of re-fitting the data, we simply calculate the mean data, best-fit model and accretion rate $\dot{m}$ by averaging the corresponding bins, weighted by their source numbers. It is clear that the disky warm Comptonization emission (green dashed line) is very different in the three bins, while the total X-ray coronal luminosity (integration of the blue dashed line) remains similar, despite a systematic change in spectral slope.

\par
\begin{figure*}
\centering
\subfloat{\includegraphics[width=0.98\textwidth]{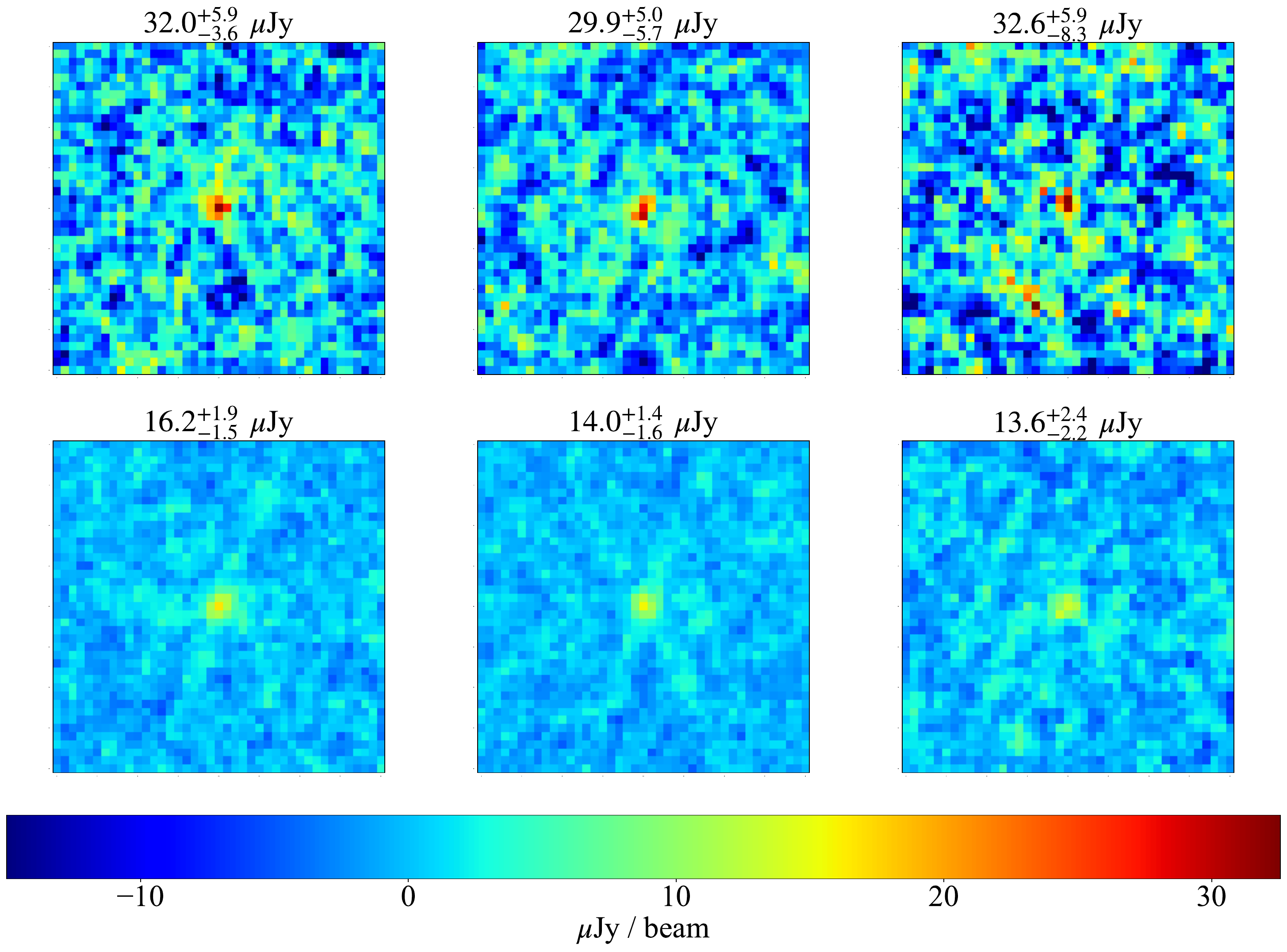}}
\caption{\label{fig:radio} Top row: stacked VLASS images of the AGN in each of the 3 SED bins: faint (left), middle (middle) and bright (right) bins. Bottom row: stacked VLASS images of the control sample of non-AGN galaxies matched in redshift, stellar mass and in star-forming properties. All the co-added images have a scale of $41\arcsec\times 41\arcsec$ with pixel size = 1\arcsec.
}
\end{figure*}

 The stacked VLASS images are shown in Figure \ref{fig:radio}. For all the three luminosity bins, we detect a signal at $> 5 \sigma$ level. Strikingly, the radio fluxes are statistically the same among the three bins, despite their vastly different optical/UV to X-ray SEDs. 

This could be explained if the radio follows the X-ray flux, rather than the total mass accretion rate, as in the stellar mass binary systems. However, the constancy of radio flux between these bins could also be explained if it is dominated by the host galaxy, as all these bins have similar host galaxy masses. To explore this, we need to estimate and subtract any host contamination before comparing the intrinsic radio luminosities.

\begin{figure*}
\centering
\subfloat{\includegraphics[width=0.99\textwidth]{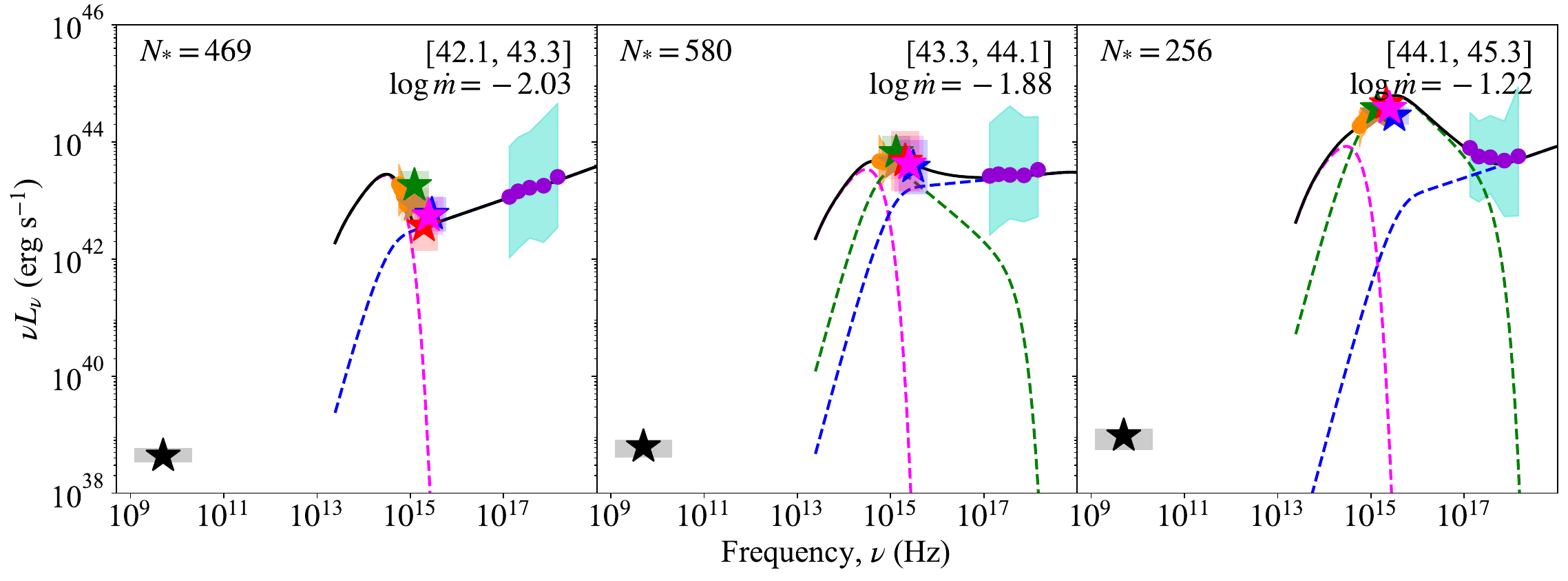}}
\caption{\label{fig:sed_radio} The stacked VLASS data (black star), along with the optical to X-ray SED in the three bins derived by averaging those in \citet{Hagen24} and \citet{Kang_2025}. Colored shades show the $1\sigma$ uncertainty of the stacked radio luminosity, or the $1\sigma$ scatter for other bands. Each panel shows the $\nu L_{3500\AA}$ luminosity bin range in the top right corner. The solid line is the best-fit \textsc{agnsed} model \citep{Kubota_2018} in \citet{Hagen24}, while the dashed lines show the individual components of the standard outer disc (magenta), warm Comptonising disc (green), and inner hot X-ray plasma (blue). See Figure \ref{fig:sed} for a detailed view of the SED. 
}
\end{figure*}  

\subsection{Radio emission of host galaxies}\label{S:host}
\par Inactive (without an AGN) galaxies are not radio-silent. They usually show broad-band radio emission produced by synchrotron radiation and free-free emission from H II regions as well as the thermal re-radiation by dust \citep[see the review by][]{Condon_1992}. In particular, a tight correlation between the far-infrared (FIR) and radio emission has been widely observed in star-forming galaxies \citep{Yun_2001, Sargent_2010}. Since the FIR emission is a direct probe of dust and cold gas, such a correlation implies the radio emission to be driven by recent star forming activities, probably synchrotron emission from supernovae shocks accelerating particles to relativistic energies. 

\par It is not yet clear whether the radio emission in radio-quiet AGN is predominantly from the AGN nucleus or from the host galaxy. Some works have shown the host galaxy dominates \citep{Bonzini_2015, Padovani_2016}, while others show an AGN-dominated \citep{Zakamska_2016,Calistro_2024} or a more complicated hybrid scenario \citep{Yue_2024}. 

\par
\begin{figure}
\centering
\subfloat{\includegraphics[width=0.48\textwidth]{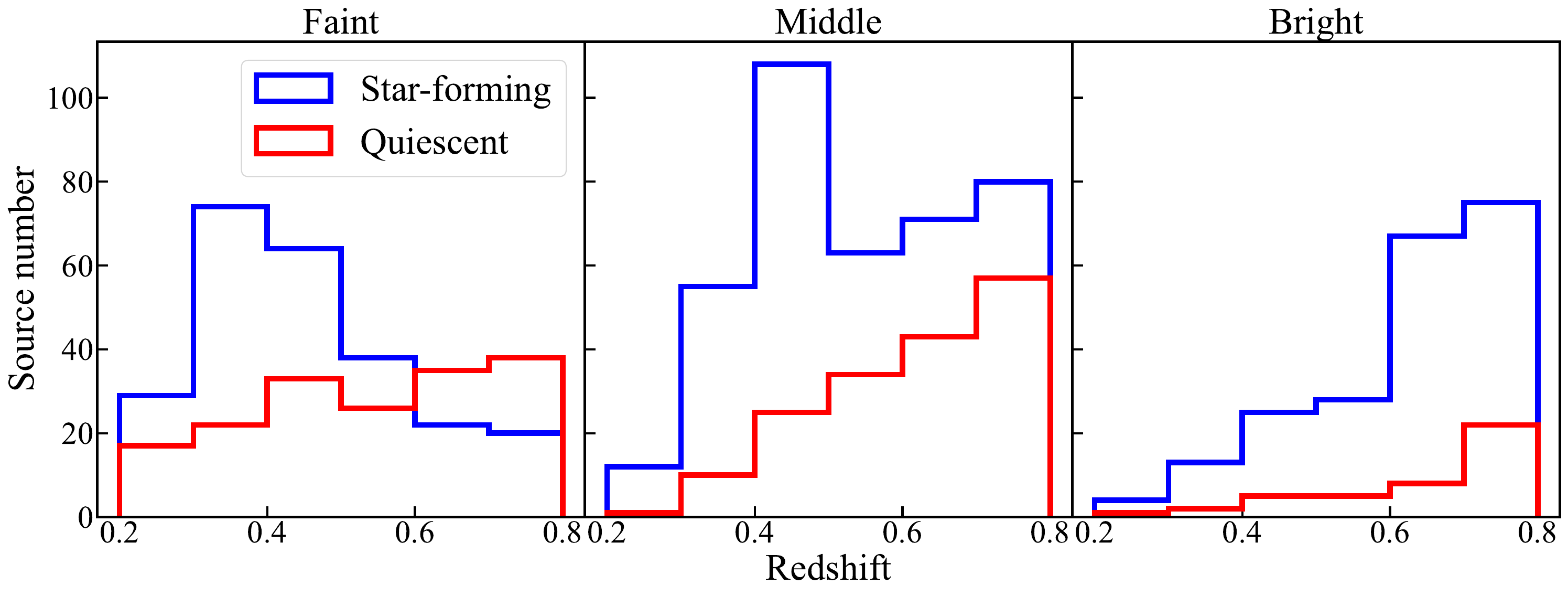}}
\caption{\label{fig:hosttype} The redshift distributions of the quiescent (red) and star-forming (blue) host galaxies in each luminosity bin. 
}
\end{figure}

 We explore this in our sample. Using the decomposed HSC images, \citet{Li_2024} has classified the host galaxies in our sample into star-forming (SF) and quiescent (QS) galaxies, based on the $u-r$ versus $r-z$ diagram. As shown in Figure \ref{fig:hosttype}, the fractions of the SF and QS galaxies are different in the three bins and are redshift dependent. It is thus difficult to estimate the contribution of the host galaxy using known empirical correlations; instead, we attempt to build a control sample of matched inactive galaxies to quantify that. 

\par
\begin{figure}
\centering
\subfloat{\includegraphics[width=0.48\textwidth]{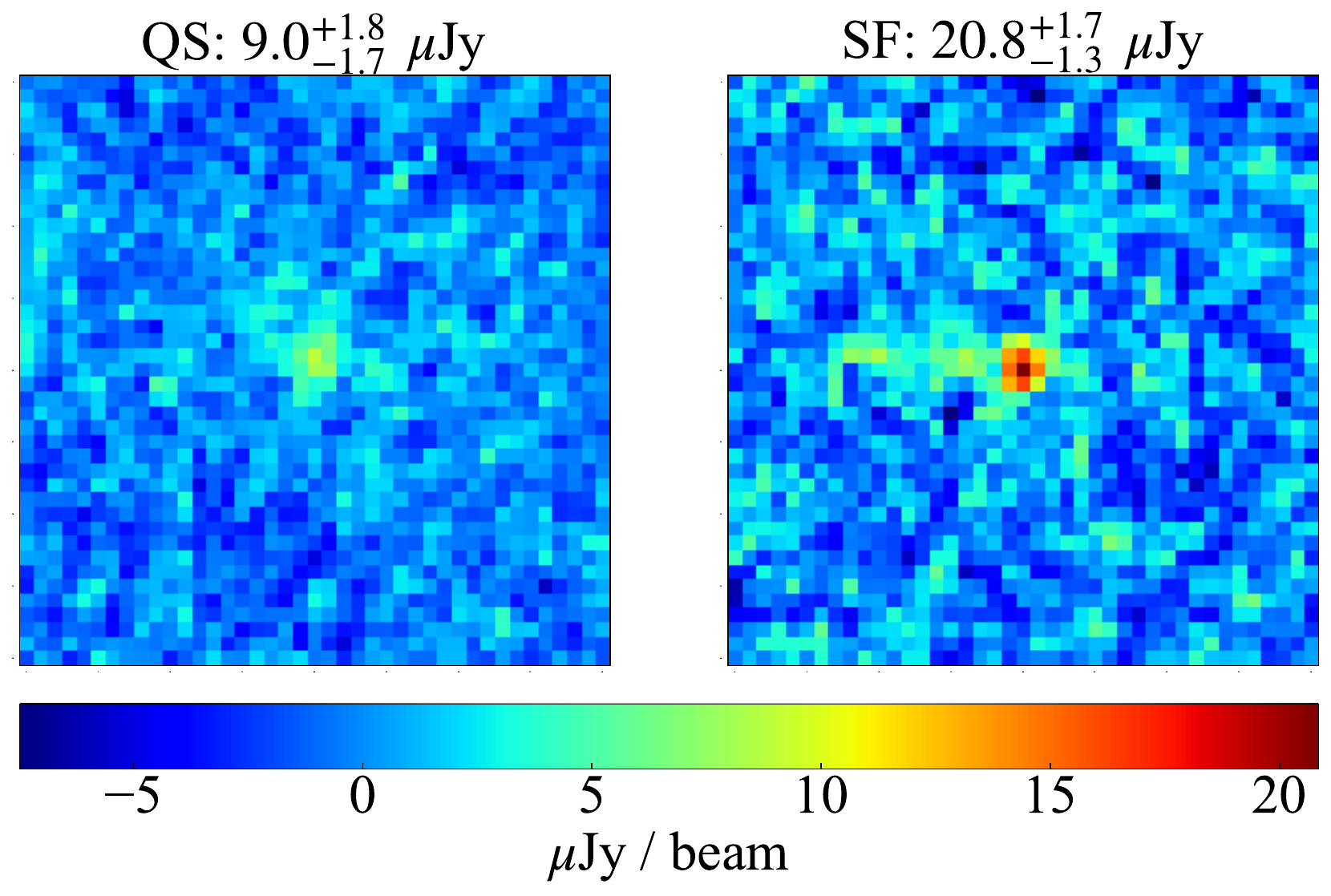}}
\caption{\label{fig:inactive} Stacked VLASS images of 4000 quiescent (left) and star-forming (right) inactive galaxies with redshift $0.4< z < 0.5$ and log $M_{\rm stellar}$ ($M_{\sun}$) [10.5--11.0]. 
}
\end{figure}

 The HSC-eFEDS sample in \citet{Kawinwanichakij_2021} (hereafter K21 sample) contains more than 1 million inactive galaxies, the stellar masses and types (SF or QS) of which have been determined in the same way as our sample \citep{Li_2024}, and thus can be used as a perfect control sample. In the first place we want to quantify the typical flux of the host galaxies in our sample. We randomly select 4000 SF and 4000 QS galaxies with log $M_{\rm stellar}$ ($M_{\sun}$) among 10.5--11.0 (typical values for our sample, and we will construct more strict control samples below), and with 
redshift evenly distributed between $z=0.4-0.5$.
We stack their VLASS images, as shown in Figure \ref{fig:inactive}. We conclude that 1) the QS galaxies in the K21 sample classifed by color-color diagram also show some radio emission, while SF galaxies are about two times brighter; 2) both types of galaxies can contribute to a significant but not dominant fraction of the observed fluxes in our AGN sample. 

\par We now build a matched control sample of inactive galaxies for our AGN sample.
For each QS/SF host galaxy in our AGN sample, we find 10 inactive QS/SF galaxies in the K21 sample with similar redshift and $M_{\rm stellar}$. Such a control sample should well represent the properties of the host galaxies in our sample, unless some key parameters (e.g., star-formation rate and size) are strongly correlated with AGN activities. The stacked VLASS images of the control sample are shown in the lower panels of Figure \ref{fig:radio}. In all the three bins, the host galaxy contributes $\lesssim 50$ percent of the observed flux. We then adopt the host-subtracted fluxes as the intrinsic AGN radio fluxes in the following analysis. These give an upper limit to any radio emitting compact jet from the AGN, as these host subtracted radio fluxes could still be contaminated by emission from either unresolved nuclear star-formation, or shocks from AGN winds. 

\subsection{Radio luminosity and the fundamental plane} \label{sec:FP}

\par
\begin{figure}
\centering
\subfloat{\includegraphics[width=0.48\textwidth]{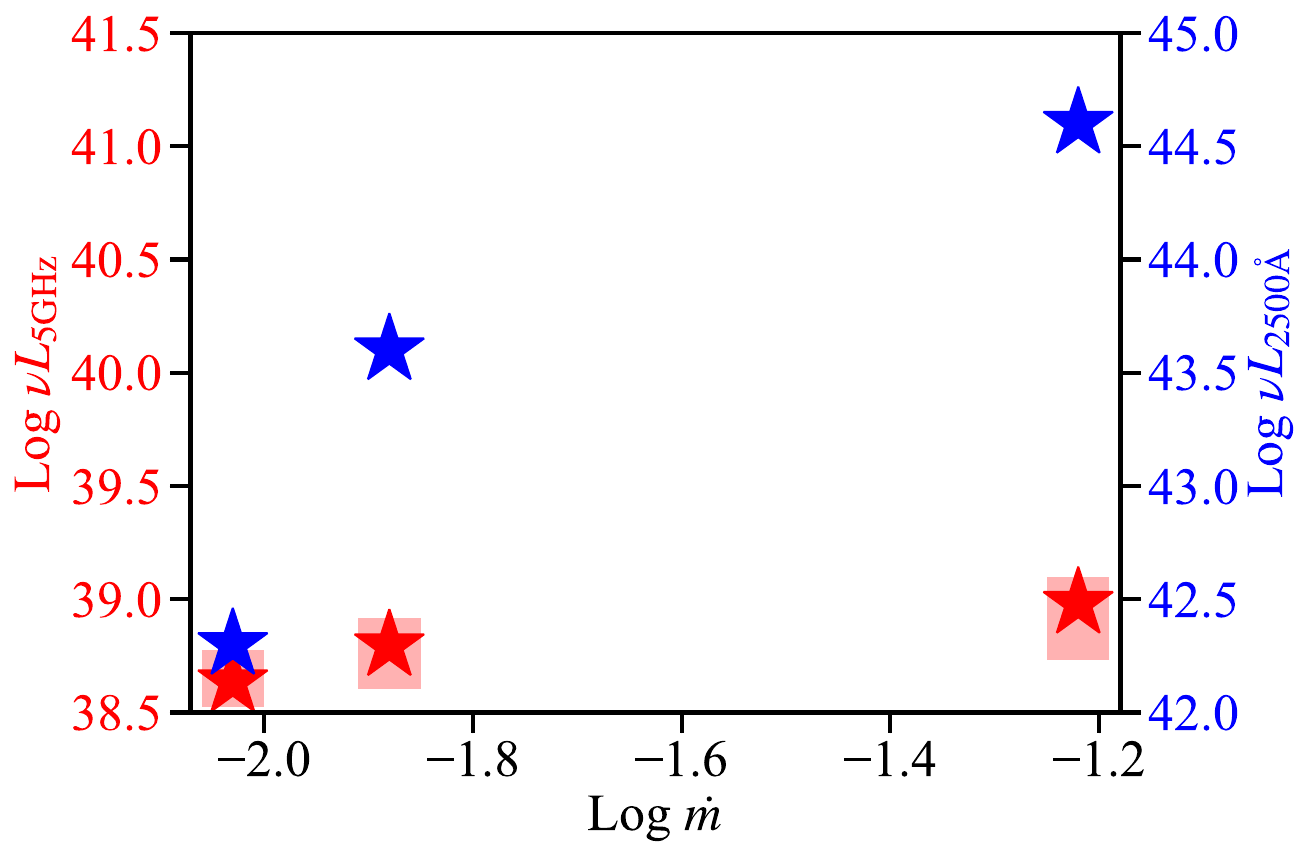}}
\caption{\label{fig:mdot_lumis} Log $\dot{m}$ versus $\nu L_{2500 \AA}$ (blue) and $\nu L_{\rm 5~GHz}$ (red) in the three bins. The UV luminosity is changing by a factor than 100 while radio changes only by a factor 2 for a factor 6 change in bolometric luminosity (i.e.  mass accretion rate).}
\end{figure}

 We convert the host-subtracted VLASS fluxes into rest-frame 5 GHz luminosities, assuming a power law spectrum with a slope $\alpha_{\nu} = -0.5$ \citep[e.g.,][]{Jiang_2007}, and distances calculated from the mean redshift in each bin using the Planck 2018 cosmology parameters \citep{Planck_2020}. The $\nu L_{\rm 5~GHz}$ of the faint, middle and bright bins are $4.3$, $6.2$ and $9.6 \times 10^{38}$ $\,\rm erg\,cm^{-2}\,s^{-1}$, respectively. The bright bin is only twice as luminous as the faint bin in radio emission, while its accretion rate is $\sim 6\times$ higher and UV emission is $\sim 100 \times$ brighter (see Figure \ref{fig:mdot_lumis}).

\par Therefore, we conclude the radio power of AGN is not dependent on the power of the accretion disk or the overall accretion rate. 
Instead, the relatively constant 
radio power is more consistent with correlating with the relatively constant X-ray power. 
Such a correlation is predicted by  
the `fundamental plane of black hole activity' \citep[e.g.,][]{Merloni_2003, Falcke_2004, Wang_2006, Yuan_2009, Bonchi_2013, Xie_2017, Bariuan_2022, Wang_2024}. This
is an empirical correlation between the X-ray and radio luminosity of stellar and supermassive black holes at varying $\dot{m}$, but has theoretical underpinning if the X-rays are assumed to be from a radiatively inefficient flow, where the accretion power $P_{\rm acc}=\dot{m}M$ but the radiated X-ray flux is 
$L_{\rm X}\propto \dot{m} P_{\rm acc}$ \citep[as the efficiency increases with $\dot{m}$,][]{Narayan94}. This connects to the base of a \citet{Blandford_1979} self-similar conical jet with power $P_{\rm jet}\propto P_{\rm acc}$ \citep{Heinz_2003}. The radio flux is given by the sum of self absorbed synchrotron components from the vertically extended, self similar jet, leading to a flat spectrum in the radio with monochromatic $L_{\rm R}\propto (\dot{m} M)^{17/12}$ \citep{blandford_1983,Heinz_2003,Merloni_2003}. Substituting for $\dot{m}$ from the  $L_{\rm X}$ relation gives a {\em prediction} that $L_{\rm R}\propto (L_{\rm X} M)^{17/24}$ so $\log L_{\rm R}=0.71\log L_{\rm X} +0.71 \log M +C$ where $C$ is a constant. 
A more detailed (post-diction!) examination of ADAF properties gives $L_{\rm X}\propto \dot{m}^{2.3}M$, matching the observed trend from the data of 
$\log L_{\rm R}=0.60\log L_{\rm X} +0.78 \log M +C$ \citep{Merloni_2003}.

\par
\begin{figure}
\centering

\subfloat{\includegraphics[width=0.48\textwidth]{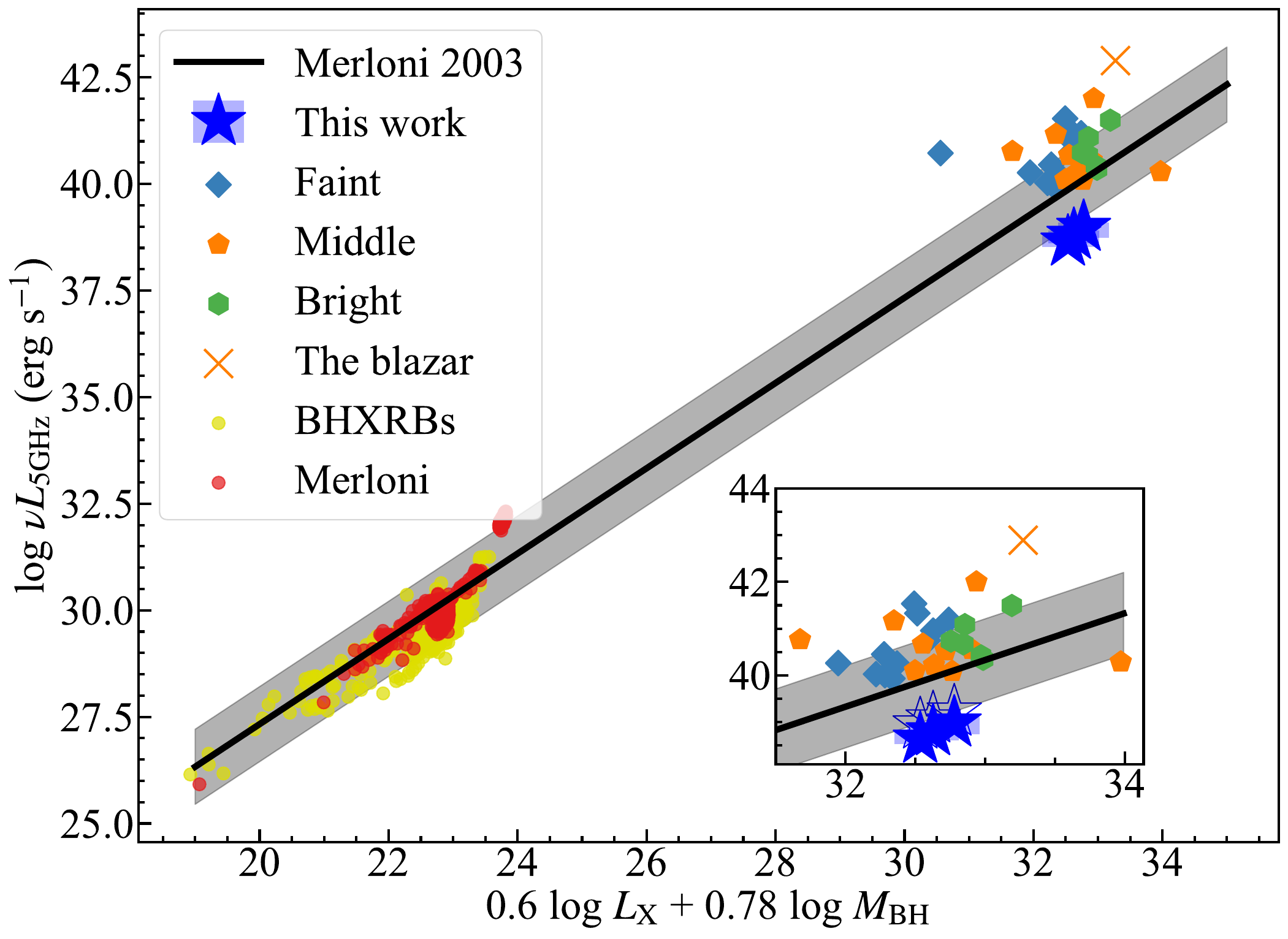}}
\caption{\label{fig:fundamental_plane} The three luminosity bins of the eFEDS-HSC sample in this work plotted on the fundamental plane of black hole activity extended down to the BHXRBs range \citep{Merloni_2003}. $L_{\rm X}$ refers to the rest-frame 2--10 keV luminosity derived by integrating the hot Comptonization component of the best-fit \textsc{agnsed} model. The inset zooms in to the AGN region for our sample, with the open stars showing the radio luminosities without the subtraction of the host galaxies. We also show the 29 individually detected AGN with colours and shapes indicating their SED bin (faint: lightblue diamand, middle: orange pentagon, bright: green hexagon). The cross marks the one known blazar in the sample. It is very clear that the individually detected objects are brighter with respect to their observed X-ray emission than the stacked undetected ones. We also include BHXRBs with available X-ray and radio data. Red points indicate those in the original sample of \citet{Merloni_2003}, while yellow are more recent detections.}

\end{figure}

\begin{figure}
\centering
\subfloat{\includegraphics[width=0.48\textwidth]{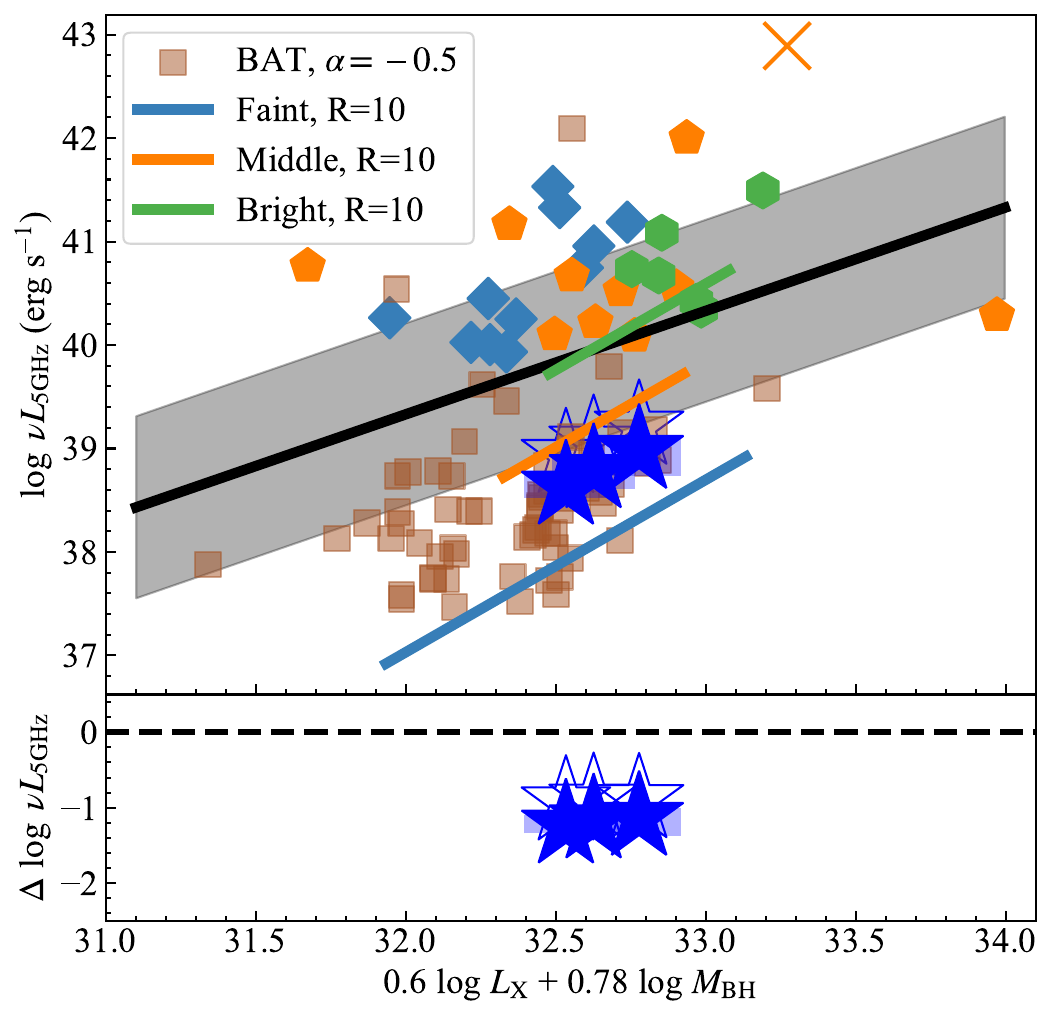}}
\caption{\label{fig:fundamental_plane_zoom} Top panel: zoom in of Figure \ref{fig:fundamental_plane} with addional hard X-ray selected AGN from the SWIFT BASS VLA sample for this fixed mass bin of $\log M_{\rm BH}/M_{\sun} =8.0-8.5$ (brown squares). These are a nearby ($z<0.05$), almost complete radio detected sample, and are observed at high spatial resolution so their radio emission is not contaminated by the host galaxy \citep{Magno_2025}. This makes them an excellent comparison set to our stacked, host galaxy subtracted points. The bulk of the BASS sample are consistent with our stacks, but again there is a tail with significantly higher radio emission for the same mass and mass accretion rates. We also mark on the classic $R=10$ RL/RQ distinction for our three SED bins with coloured solid lines. This shows that our stacks, and the bulk of the BASS AGN, will change from RL to RQ at the spectral transition, not due to any change in the jet but solely driven by the change in disc. Distance away from the FP relation (radio to X-ray ratio) is a much better indicator of the intrinsic jet power than the classic radio loudness (radio to optical/UV ratio). Bottom panel: the deviations of our stacked points from the \citet{Merloni_2003} FP relation. $\Delta \log \nu L_{\rm 5GHz}$ are statistically the same for the three luminosity bins, which are $\sim$ 1.2/0.9 dex, for the stacked data with/without host subtracted respectively. 
}
\end{figure}

 First we simply plot our three bins on the empirical FP relation derived by \citet{Merloni_2003}. They define $L_{\rm X}$ from the 2--10 keV X-ray luminosity of the hot corona so we integrate our AGN 
best-fit hot Compontonized model (blue dashed line in Figure \ref{fig:sed_radio}) over this range. This gives log $L_{\rm X}$ ($\,\rm erg\,s^{-1}$) of 43.51, 43.64 and 43.88, for the faint, middle and bright bin, respectively. 
We combine this with the radio luminosities in each bin to plot our 3 stacked SEDs on the FP relation in Figure \ref{fig:fundamental_plane}. Our three bins are below (radio fainter) by $\sim$ 1.2 dex than predicted by the empirical relationship, and the decrepancy is statistically the same in the three luminosity bins (see the bottom panel of Figure \ref{fig:fundamental_plane_zoom}).


We note the consistency would be slightly better (despite still a decrepancy of 0.9 dex)
if we had plotted the observed radio fluxes (open stars, see inset in Figure \ref{fig:fundamental_plane} and \ref{fig:fundamental_plane_zoom}), rather than correcting for the contribution of the host galaxy and indeed \citet{Merloni_2003} do not subtract any host galaxy contribution. 
However, more importantly, \citet{Merloni_2003} only use objects which are {\it detected} in radio flux, so are likely biased high relative to our sample. 
We test this by showing the 29 individually detected VLASS sources separately on Figure \ref{fig:fundamental_plane}, colour coding to distinguish between those in the faint (lightblue diamand), middle (orange pentagon) and bright (green hexagon) bins \footnote{We use a simple power law with Galactic absorption corrected to fit their eFEDS spectra and calculate the 2--10 keV $L_{\rm X}$. The eROSITA spectra of the several faintest sources are not good enough for a robust fitting (see last few panels of Figure \ref{fig:indi_SED}), which cause underestimation of their $L_{\rm X}$. The obvious outlier (lightblue diamand with extremly low $L_{\rm X}$) in Figure \ref{fig:fundamental_plane} is source 31206, shown in the last panel of Figure \ref{fig:indi_SED}.}.
There is only one known blazar in the sample (marked with a cross in Figure \ref{fig:fundamental_plane}), where the radio is known to be strongly enhanced by Doppler boosting, but all the other individually detected sources
are systematically above the \citet{Merloni_2003} relation, so it is clear that radio flux limited samples will bias high when the radio detections are incomplete. In other words, \citet{Merloni_2003} has likely selected AGN with intrinsically higher radio jet efficiency compared with the whole population.

\par We also include data from black hole X-ray binaries, using the `Radio/X-ray correlation database for X-ray binaries' \citep{arash_bahramian_2022_7059313}. We assume $M_{\rm BH} = 10 M_{\odot}$ and a power law X-ray spectrum with $\Gamma = 2.0$. We find these binaries, many of which were studied after \citet{Merloni_2003}, are also slightly radio fainter on average than the original FP. We highlight the original sources in \citet{Merloni_2003} in red and see that the BHXRBs population is also slightly below the original FP of \citet{Merloni_2003}.
Therefore, the constant in the FP of \citet{Merloni_2003} is likely overestimated. 

We explore this further using the (58-month) Swift BAT ultra-hard X-ray flux limited sample of AGN \citep{Mushotzky_2014}, where the $z<0.05$ subset in the North has recently been surveyed at high spatial resolution and high sensitivity with VLA imaging at 22~GHz \citep{Magno_2025}. This is an almost complete radio detected subsample, so should be relatively unbiased. We select the AGN with masses in our range of $\log M_{\rm BH}/M_{\sun} =8.0-8.5$, extrapolate the 22~GHz flux of the core (1\arcsec~flux) down to 5~GHz again assuming a power law $F_\nu\propto \nu^{\alpha}$ with $\alpha=-0.5$. We show these as brown squares on Figure \ref{fig:fundamental_plane_zoom}. There are only 2 sources above the FP (PKS2331-240, which is a Fermi detected misaligned jetted AGN in a giant radio galaxy: \citealt{Paliya_2025}, and Arp102B), the rest are below, but now with individually detected sources rather than a stack. These Swift BAT AGN pick out the same $L_{\rm R}-L_{\rm X}$ relation as our 3 stacked spectra in the same range of $L_{\rm X}$. This gives confidence in the level of radio detected from our stacked sample, as the VLA resolves out all large scale extended star formation in this very low $z$ sample. However, there can still be small scale extended radio emission contaminating the expected compact jet, and higher resolution VLBA images are clearly a better way to explore this \citep{Fischer_2021,Chen_2023}. The Swift BAT AGN extend down to lower $L_{\rm X}$ than our stacks as these are 
all relatively nearby objects, so lower $\dot{m}$ objects can be detected. 

This dispersion in radio flux {\em on the FP} (i.e. corrected for X-ray luminosity) is more than a factor 100 for AGN of the same mass and mass accretion rate. Clearly there is at least one other parameter which is important in setting the radio luminosity. Jet power is known to depend strongly on black hole spin, 
via the \citet{Blandford77} process \citep{blandford_1983, Tchekhovskoy_2010}, so this could be a major source of this dispersion, though the jet power will also depend on the 
history of accumulation of magnetic flux pinned onto the black hole \citep{Begelman_2012}, and/or there may be a 
supermassive counterpart of the transient discrete ballistic jets which are triggered at the spectral transition in BHXRBs \citep{Nipoti_2005}.

These additional processes can enhance the power of the radio emission above that predicted by the FP extended from the stellar mass BHXRBs. The BHXRBs discrete ejections are already excluded from the radio-X-ray correlation, so extrapolating from these sources to the AGN depends on the mean BHXRBs spin and magnetic flux. Spin can be measured in BHXRBs but this is currently controversial 
\citep{Belczynski_2024, Zdziarski_2024b, Zdziarski_2025}. Here we assume they have mean spin of around $0.5-0.7$ as indicated from fitting a sequence of disc spectra of fixed innermost stable orbit to the disc dominated spectra \citep{Gierlinski_2004, Done_2007}. AGN spins (and masses) depend on their history of accretion over cosmic time, so they can have a wide range in spin \citep{Volonteri_2005, Beckmann_2025}. We assume no net magnetic flux, i.e. the accretion flow is SANE \citep{Zdziarski_2024}. This predicts that the AGN which are lower relative to the FP are then lower spin than the BHXRBs, while those above have higher spin. \citet{Unal_2020} show that the Blandford-Zdnajek process can explain a factor $>100\times$ more jet power at high spin ($a>0.9$) than at moderate spins ($a\sim 0.5-0.6$), as is also seen in numerical simulations \citep{Tchekhovskoy_2010}.

\subsection{Fundamental plane versus radio-loudness}

\par The stacked radio from our sample is more or less consistent with an origin in a compact steady jet, which scales in a self similar way with the X-ray hot flow across the mass scale from BHXRBs as predicted by the FP \citep{Merloni_2003}. Our AGN sample spans above and below the very obvious spectral transition at $L\sim 0.01L_{\rm Edd}$, and we find that these are consistent with the same disc-jet coupling, with very little change in either the radio or X-ray luminosity despite a large change (factor 6) in mass accretion rate and an even larger change (factor 100) in UV flux. However, this large change in disc flux does give a large change in the radio to optical/UV ratio which is the classic definition of radio jet power via the radio-loudness parameter. 

We define the radio-loudness parameter as $R=L_{\rm 5~GHz}/L_{\rm 2500\AA}$. We measure $L_{\rm 2500\AA}$ in each of our 3 stacked spectra (Figure \ref{fig:sed_radio}) and determine $L_{\rm 5~GHz}$ for which $R=10$. 
These lines are shown in Figure \ref{fig:fundamental_plane_zoom} for the faint (lightblue), middle (orange) and bright (green) bins. 
The bin width in black hole mass and X-ray luminosity give the extent of the RL/RQ line for the bright and middle bins, but the lowest luminosity SED is extended down to much lower $L_x$ due to the inclusion of lower $\dot{m}$ sources in the BASS sample.

Our stacked radio from the undetected sources and the bulk of the BASS AGN radio sample for black holes of the same mass lie between the $R=10$ lines for the faint and middle sample. Thus the majority of these AGN will change classification at the spectral transition from RL (faint) to RQ (middle-bright), yet the radio shows almost no change \citep[see also][]{Terashima_2003}. The change in designation is driven solely by the change in disc component not the jet itself.

This all supports the discussion above where distance from the FP is taken as a better measure of relative jet power. However, we notice that the $R=10$ line for bright AGN coincides with the FP for this AGN mass range, illustrating why this single number has been so useful in identifying high power jets. However, it also has clear drawbacks, firstly as illustrated here at the spectral transition, and secondly as it does not correct for the increasing radio dominance expected 
for the same self absorbed synchrotron spectrum at higher masses \citep{Schulze_2017}.

\subsection{Other origins of the scatter on the FP}

\par Besides the black hole spin, several factors could also contribute to the scatter on the FP. Here we use a sample of AGN with $\log M_{\rm BH}/M_{\sun} =8.0-8.5$, where $M_{\rm BH}$ are estimated from the HSC decomposition and the empirical relation between the stellar and black mass ($M_{\star}$ and $M_{\rm BH}$). The main uncertainty of these masses comes from the empirical relation since the HSC decomposition is believed to be rather reliable, and the typical uncertainty level is $\sim 0.4$ dex \citep{Ding_2020}. Such uncertainty of the mass measurement could induce some scatter on the FP as well as some systematics across the luminosity bins. For example, brighter sources are more likely to be heavier AGN fluctuated into the $8.0-8.5$ bin, i.e., have underestimated masses.

\par Additionally, galaxy morphology and  environment differ in different sources, which could influence the radio emission efficiency and cause scatter on the FP. However, our mass range of $\log M_{\rm BH}/M_{\sun} =8.0-8.5$ is directly taken from the host galaxy stellar mass range of $\log M_{*}/M_{\sun} =10.74-11.25$. The limited stellar mass range means there is a limited range in morphology and environment. This is not true for samples extending over a much larger range in galaxy masses. In particular, the highest mass galaxies (and highest mass black holes) tend to be Central Dominant elliptical galaxies, in a highly clustered environment. Our mass range avoids these dense environments, so gives a more homogeneous sample. 

\par AGN variability could also influence the observed radio-X-ray relation. 
Non-blazar AGN generally show significant X-ray variability on timescales from $\sim$ 100s to years \citep[e.g.,][]{Middei_2017, Hagen_24b}, while radio variations have been observed on a timescale of months with typical RMS of 10\% to 20\% \citep{Falcke_2001, Mundell_2009}. For our sample, the X-ray and radio data are only partially overlapping. The eFEDS observations were conducted in 2019-11, while the three epoch VLASS observations of the eFEDS field were conducted around 2018-01, 2020-08 and 2023-03, respectively. This would add to some scatter on the FP relation for individually detected AGN, but by stacking sources we should have significantly reduced the influence of stochastic variations. We also note that the X-ray and radio data adopted to derive the FP relation were not simultaneous either. However, for our X-ray selected sample \citep[eFEDS counts $> 10$,][]{Hagen24}, we do tend to select sources in their X-ray brighter state, leading to the stacked X-ray luminosity slightly overestimated.  

\par In summary, all these effects would contribute to some scatter on the FP relation, but seem insufficient to account for the observed large scatter of $\gtrsim$ three orders of magnitude. Instead, the distribution of the black hole spin which strongly affects the jet power \citep{Blandford77,Tchekhovskoy_2010,Unal_2020} should be the dominant origin of the observed dispersion.

\section{Discussion}

We show above that the clear spectral transition seen in the stacked eROSITA 
sample does not impact on the the core radio emission despite the factor 6 change in $\dot{m}$ and factor 100 change in optical/UV flux. The AGN remain on the FP, as their X-ray flux is similarly unaffected by the transition. 

This highlights a difference between the AGN and BHXRBs in terms of their X-ray emission above the transition. In BHXRBs, the X-ray coronal flux above the transition can be extremely weak, and current radio flux sensitivity limits make it difficult to follow. Cyg X-1 is currently the only BHXRBs which can be reliably tracked across the transition due to its distinctive, quite strong X-ray coronal emission in its soft state and its relatively nearby distance. The radio drops dramatically across the transition relative to the lower energy X-ray bandpass where the disc dominates, but this is due to the increase in disc flux \citep{Zdziarski_2011}. Instead, the ratio of radio to {\em coronal} (non-disc) X-ray flux remains on the FP but both radio and coronal X-ray fluxes are lower than in the hard state \citep{Zdziarski_2024}. A few other BHXRBs have much weaker constraints for their soft state radio and coronal X-ray emission, and these may imply that the radio emission is weaker than predicted by the FP (LMC X-1 and LMC X-3 \citealt{Gallo_2003}, GX339-4: \citealt{Corbel_2013}). Nonetheless, even in Cyg X-1, the radio and coronal X-rays both drop markedly at the spectral transition \citep{Zdziarski_2020}, whereas both remain roughly constant in our AGN sample across the same transition. Coronal X-rays are ubiquitous in bright AGN \citep{Brandt_2015}, 
unlike the BHXRBs where the disc can completely dominate the luminosity output above the spectral transition. 

We speculate that this difference in X-ray (and radio) properties between the stellar and supermassive black holes comes from their very different environments. Stellar mass black holes with low and intermediate mass companions accrete via Roche lobe overflow, so the material has high angular momentum. By contrast, in AGN (and in wind fed high mass X-ray binaries) there can more easily be a larger scale height, lower angular momentum accretion flow as well as a high angular momentum disc. Cyg X-1 may lie in between these two extremes, as it is a high mass binary with a strong stellar wind but almost completely fills its Roche lobe. If the coronal X-ray emission in AGN after the transition is always from a large scale height flow at approximately the maximum ADAF luminosity \citep{Kubota_2018,Mitchell_2023}, then the FP correlation is maintained even after the disc formation, as seen here. 
The FP models are then not only applicable to low luminosity, hard state flows
which are radiatively inefficient. They can also extend to bright AGN states above the transition. 

\section{Summary and Conclusions}

\par We recently demonstrated that AGN show a similar state transition to 
the stellar mass BHXRBs using stacked SEDs built from the combination of the 
eFEDS-HSC. This is a sample of X-ray selected, unobscured AGN with host galaxy subtraction from HSC imaging. Selecting only those with similar host galaxy masses, and using this to determine black hole mass, gave a sequence of uniformly selected AGN spectra with $\dot{m}\sim 0.01$ to 
$\dot{m}\sim 0.3$. This clearly shows a dramatic increase in the optically thick disc component in the UV across the transition at $\sim 0.01L_{\rm Edd}$. 

Here we investigate how the compact radio jet responds to this transition 
by stacking the VLASS radio images of our eFEDS-HSC sample. The flux limits on our sample mean that any individually detected source is classed as RL, while those with only upper limits are likely RQ. We only have 29 detected sources out of a total of 1305 AGN, so we remove these and stack the remaining undetected ones, reducing to three bins of $\dot{m}$ to obtain enough signal-to-noise. 
We find that each bin is significantly detected in radio emission at approximately similar flux levels. We correct for extended radio emission from the host galaxies by subtracting a matched sample of non-AGN, matched also in the ratio of quiescent and starforming galaxies with the same stellar mass. This is responsible for only around half of the detected emission. The remaining radio 
flux remains remarkably constant across the three bins, despite these spanning the spectral transition, with mass accretion rate increasing by a factor 6 and optical/UV increasing by a factor 100. However, the X-ray flux is likewise remarkably constant across the three SED bins, and the radio/X-ray ratio of all three are consistent with lying on the FP, i.e. consistent with the expected scaling up in mass of the compact, steady jet seen in the BHXRBs. This is confirmed in a very different AGN sample of ultra-hard X-ray selected SWIFT BAT AGN. We use only the 
AGN in our mass range of $\log M_{\rm BH}/M_{\sun} =8.0-8.5$ and find that these individual, nearby, extremely well resolved AGN have radio/X-ray points which 
lie on top of the ratios derived from our stacked spectra.

This shows that both the stellar mass and supermassive black holes show similar disc-jet coupling, or rather X-ray flow-jet coupling. Both systems have radio which follows the optically thin, X-ray hot coronal flux and not the optically thick, cool disc component. 
However, it also highlights a difference which is that in BHXRBs the disc can almost completely dominate above the transition, with very little coronal flux, whereas in AGN above the transition the X-ray corona is always present with luminosity which seems to be quite close to the maximal ADAF luminosity. We speculate that this is due to a separate large scale height flow in AGN from lower angular momentum material which is not present in the Roche lobe overflow BHXRBs.

Our data also highlight the well known dispersion in radio properties of AGN, with a factor 100 difference in radio luminosity even on the FP. We highlight how the classical radio-loudness parameter (ratio of radio to optical/UV emission) is not a good tracer of the jet emission at the transition as $R$ increases due to the drop in optical disc flux, changing from RQ to RL for no change in radio jet flux. We suggest distance from the fundamental plane (ratio of observed radio flux to that predicted by the FP) is a better measure of jet power, and speculate that this is a measure of black hole spin.

\section*{Acknowledgements}

\par We thank the anonymous referee for highly contructive comments which help improve the manuscript. 
We thank Andrzej A. Zdziarski, Leah Morabito and Macon Magno for useful discussions on this paper. JK gratefully acknowledges the scholarship of University of Science and Technology of China for a visiting program to Durham University. JK and JW acknowledges support from the National Natural Science Foundation of China (grant Nos. 123B2042). SH acknowledges support from the Japan Society for the Promotion of Science (JSPS) through the short-term fellowship PE23722 and from the Science and Technologies Facilities Council (STFC) through the studentship grant ST/W507428/1. CD acknowledges support from STFC through grant ST/T000244/1 and Kavli IPMU, University of Tokyo. Kavli IPMU was established by World Premier International Research Center Initiative (WPI), MEXT, Japan. JS is supported by JSPS KAKENHI (23K22533) and the World Premier International Research Center Initiative (WPI), MEXT, Japan. This work was supported by JSPS Core-to-Core Program (grant number: JPJSCCA20210003). MJT is supported by STFC through grant ST/X001075/1.

\par This research has made use of the CIRADA cutout service at URL \url{cutouts.cirada.ca}, operated by the Canadian Initiative for Radio Astronomy Data Analysis (CIRADA). CIRADA is funded by a grant from the Canada Foundation for Innovation 2017 Innovation Fund (Project 35999), as well as by the Provinces of Ontario, British Columbia, Alberta, Manitoba and Quebec, in collaboration with the National Research Council of Canada, the US National Radio Astronomy Observatory and Australia’s Commonwealth Scientific and Industrial Research Organisation.

\par This work is based on data from eROSITA, the soft X-ray instrument aboard SRG, a joint Russian-German science mission supported by the Russian Space Agency (Roskosmos), in the interests of the Russian Academy of Sciences represented by its Space Research Institute (IKI), and the Deutsches Zentrum für Luft- und Raumfahrt (DLR). The SRG spacecraft was built by Lavochkin Association (NPOL) and its subcontractors, and is operated by NPOL with support from the Max Planck Institute for Extraterrestrial Physics (MPE). The development and construction of the eROSITA X-ray instrument was led by MPE, with contributions from the Dr. Karl Remeis Observatory Bamberg \& ECAP (FAU Erlangen-Nuernberg), the University of Hamburg Observatory, the Leibniz Institute for Astrophysics Potsdam (AIP), and the Institute for Astronomy and Astrophysics of the University of Tübingen, with the support of DLR and the Max Planck Society. The Argelander Institute for Astronomy of the University of Bonn and the Ludwig Maximilians Universität Munich also participated in the science preparation for eROSITA. The authors gratefully acknowledge Andrea Merloni's contribution to the eROSITA project in general and this sample in particular. This research has made use of the VizieR catalogue access tool, CDS, Strasbourg, France.


\section*{Data Availability}
\par The eFEDS-HSC sample is from \citet{Hagen24}, and the inactive galaxy sample of \citet{Kawinwanichakij_2021} is available at \url{https://member.ipmu.jp/john.silverman/HSC/}. VLASS images are available at \url{https://archive-new.nrao.edu/vlass/quicklook/} and cutouts at \url{cutouts.cirada.ca}.



\bibliographystyle{mnras}
\bibliography{example} 

\appendix

\section{SED and VLASS images of individual sources}

\begin{figure*}
\centering
\subfloat{\includegraphics[width=0.99\textwidth]{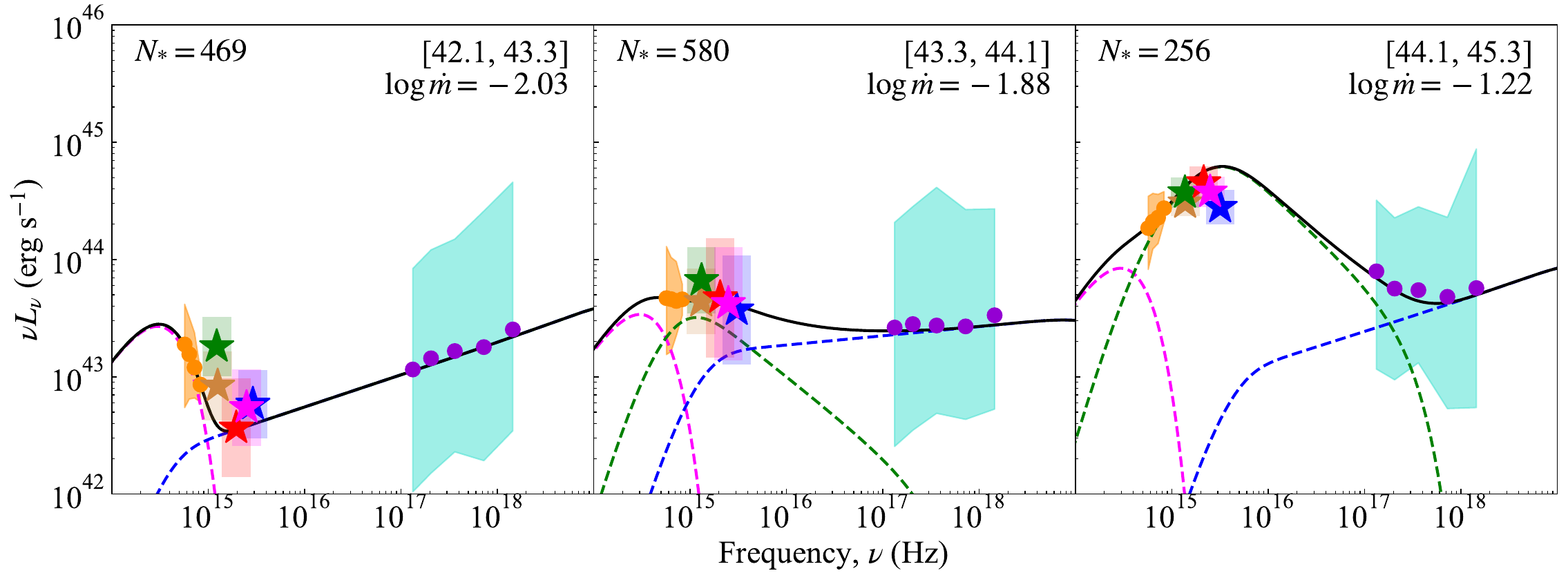}}
\caption{\label{fig:sed} The stacked optical to X-ray SED in the three bins derived by averaging those in \citet{Hagen24} and \citet{Kang_2025}. The SED points are from low to high frequency: HSC (orange points), SDSS $u$-band (green star), KiDS $u$-band (brown star), GALEX NUV (red star), GALEX 1200\AA\,(magenta star), GALEX FUV (blue star) and eROSITA (purple points). See \citet{Hagen24} and \citet{Kang_2025} for additional details.  
}
\end{figure*}

\begin{figure*}
\centering
\subfloat{\includegraphics[width=0.33\textwidth]{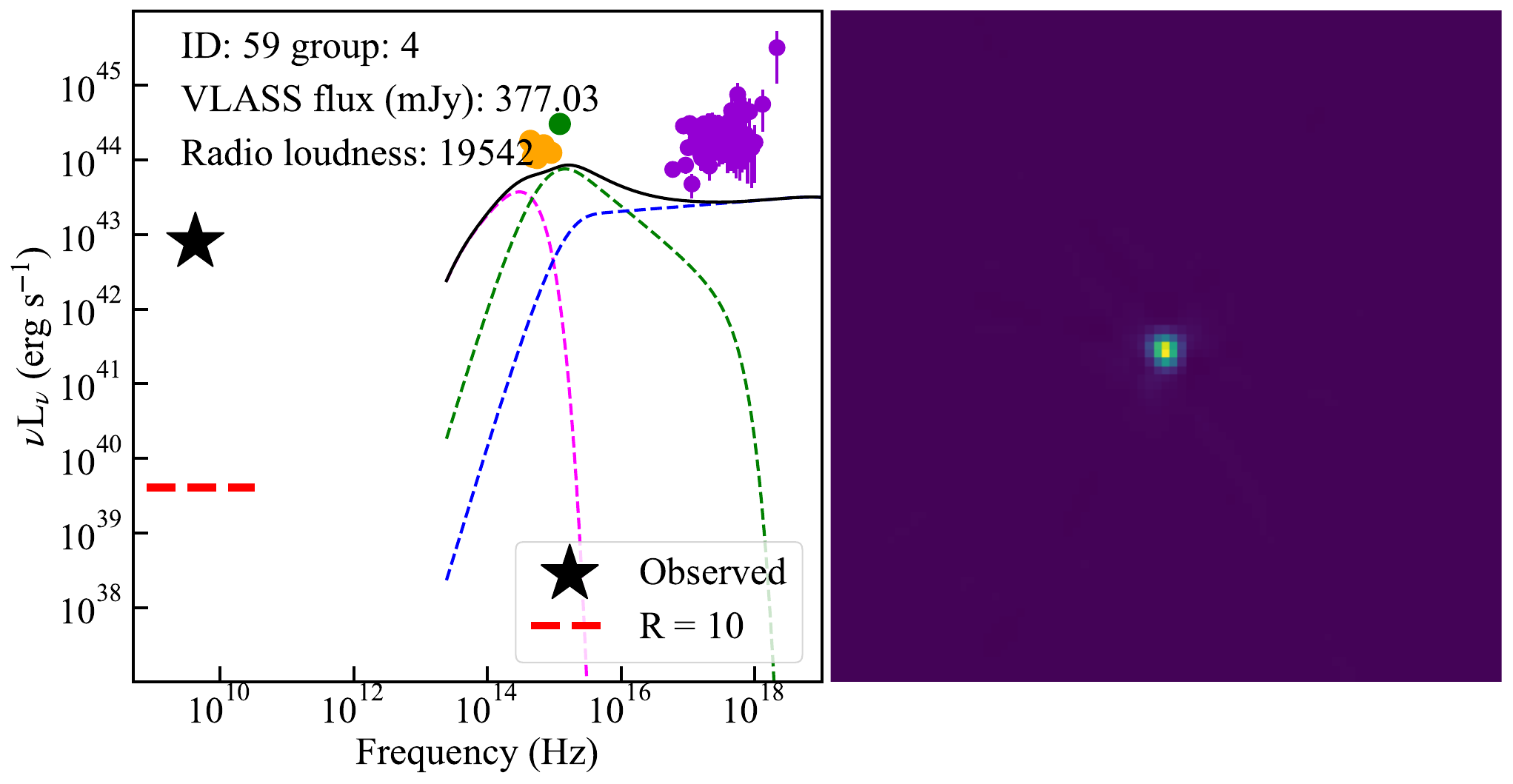}}
\subfloat{\includegraphics[width=0.33\textwidth]{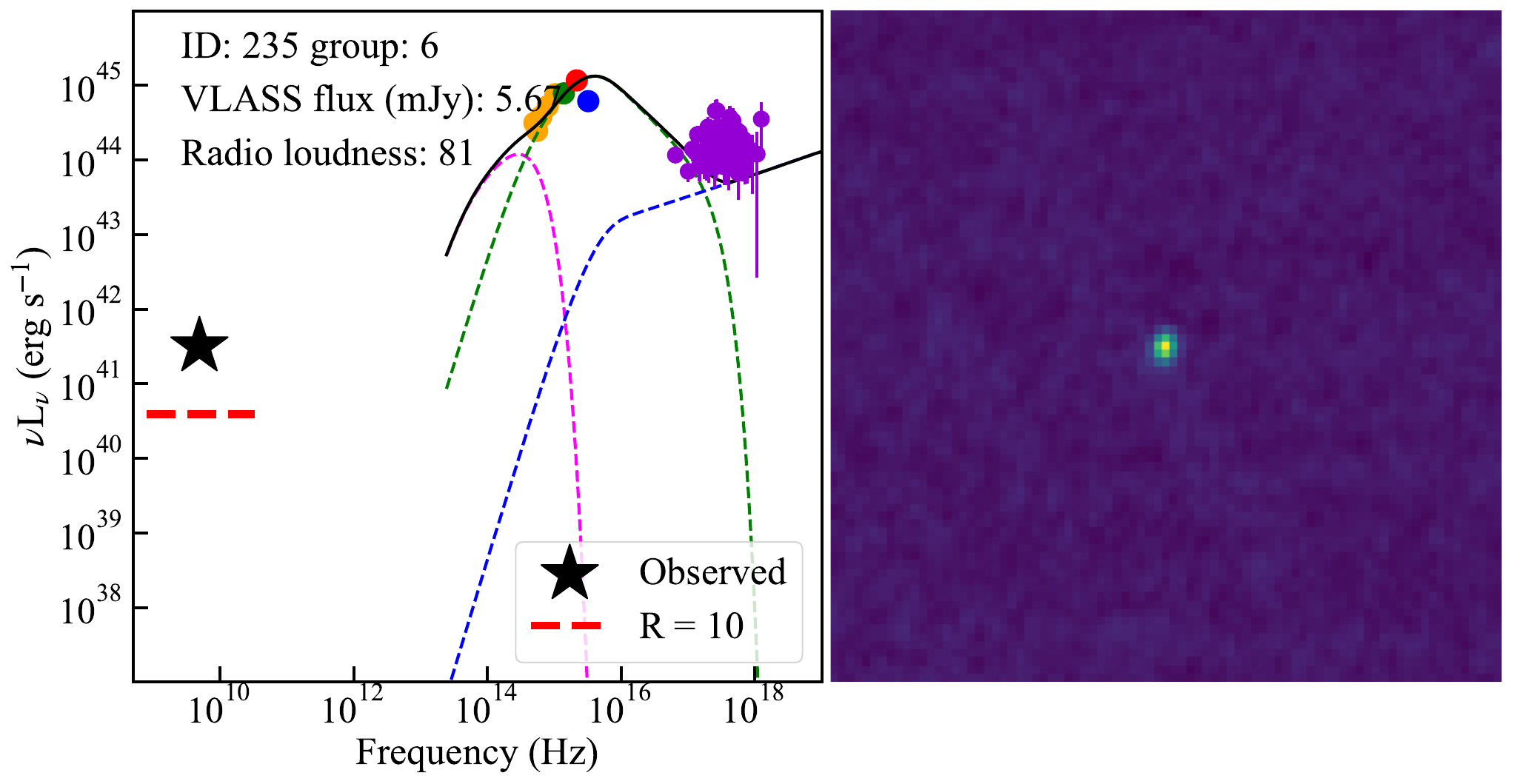}}
\subfloat{\includegraphics[width=0.33\textwidth]{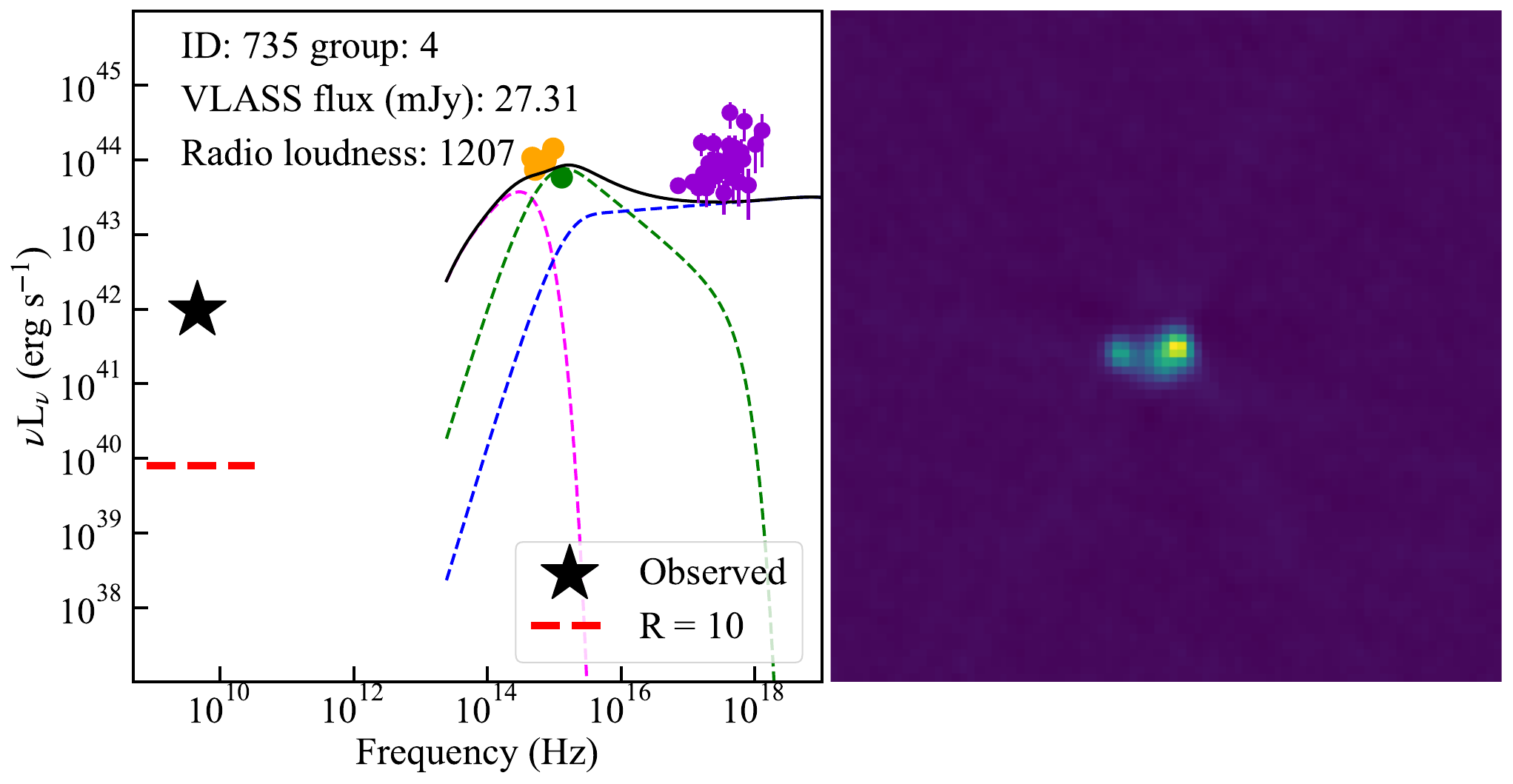}}\\
\subfloat{\includegraphics[width=0.33\textwidth]{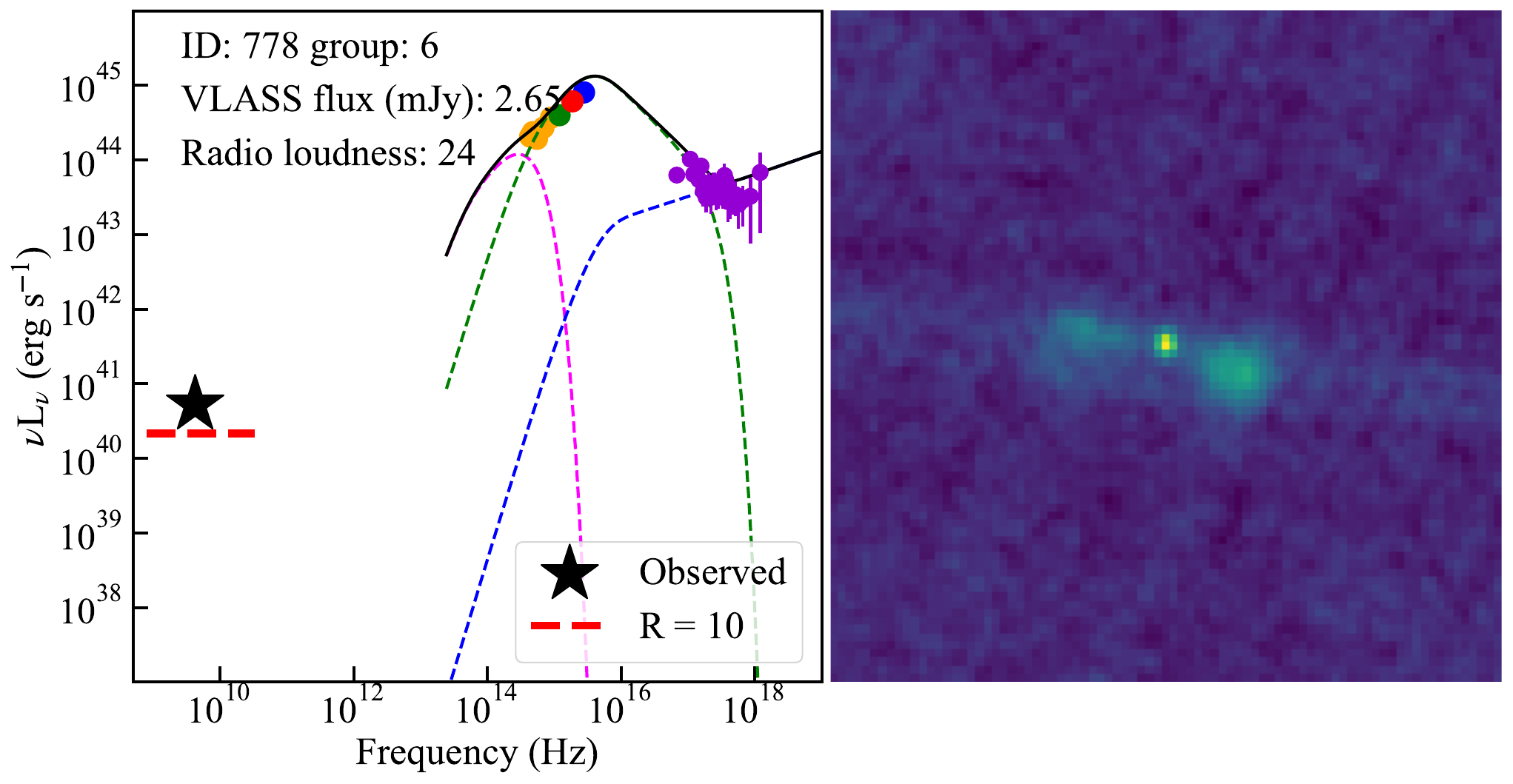}}
\subfloat{\includegraphics[width=0.33\textwidth]{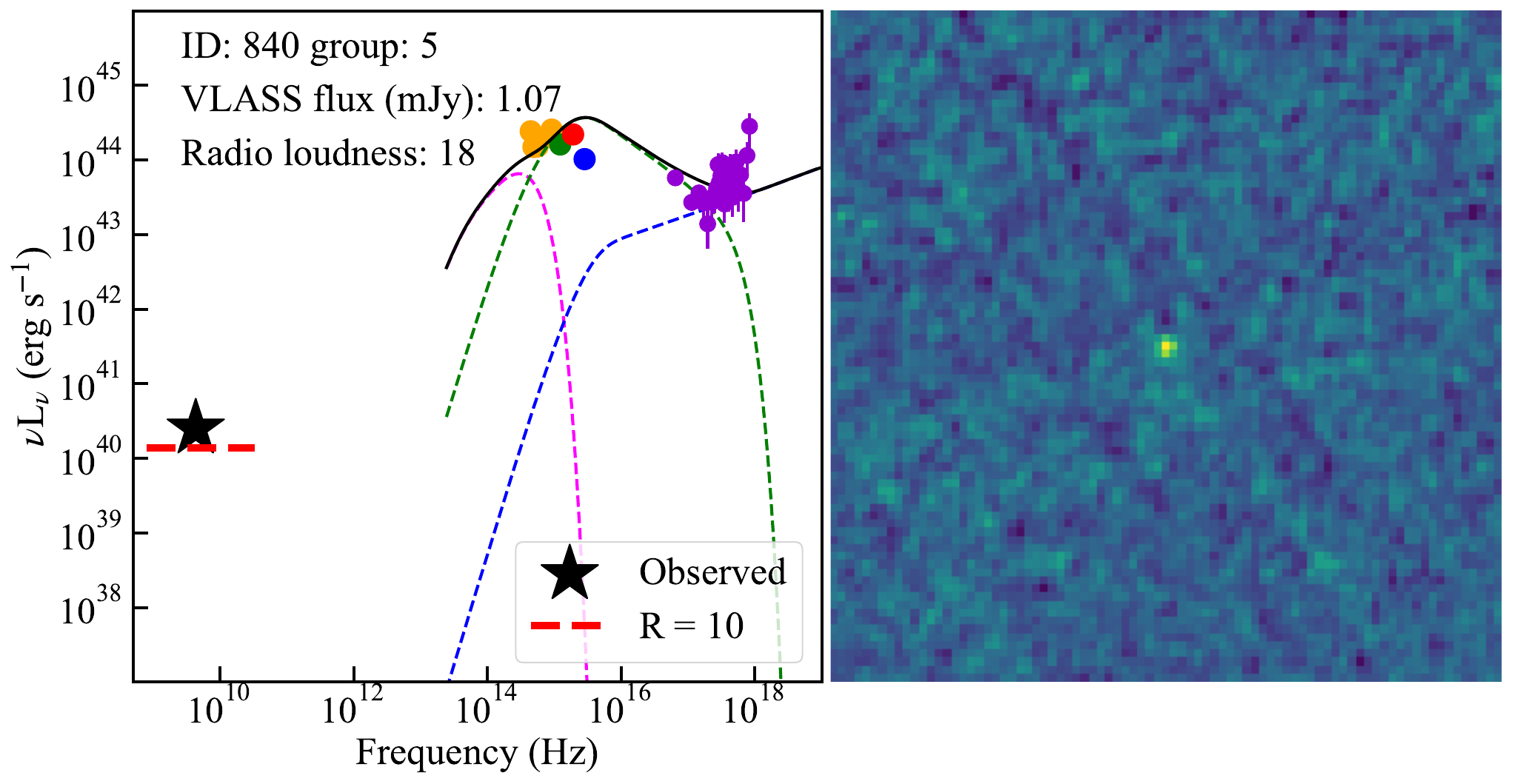}}
\subfloat{\includegraphics[width=0.33\textwidth]{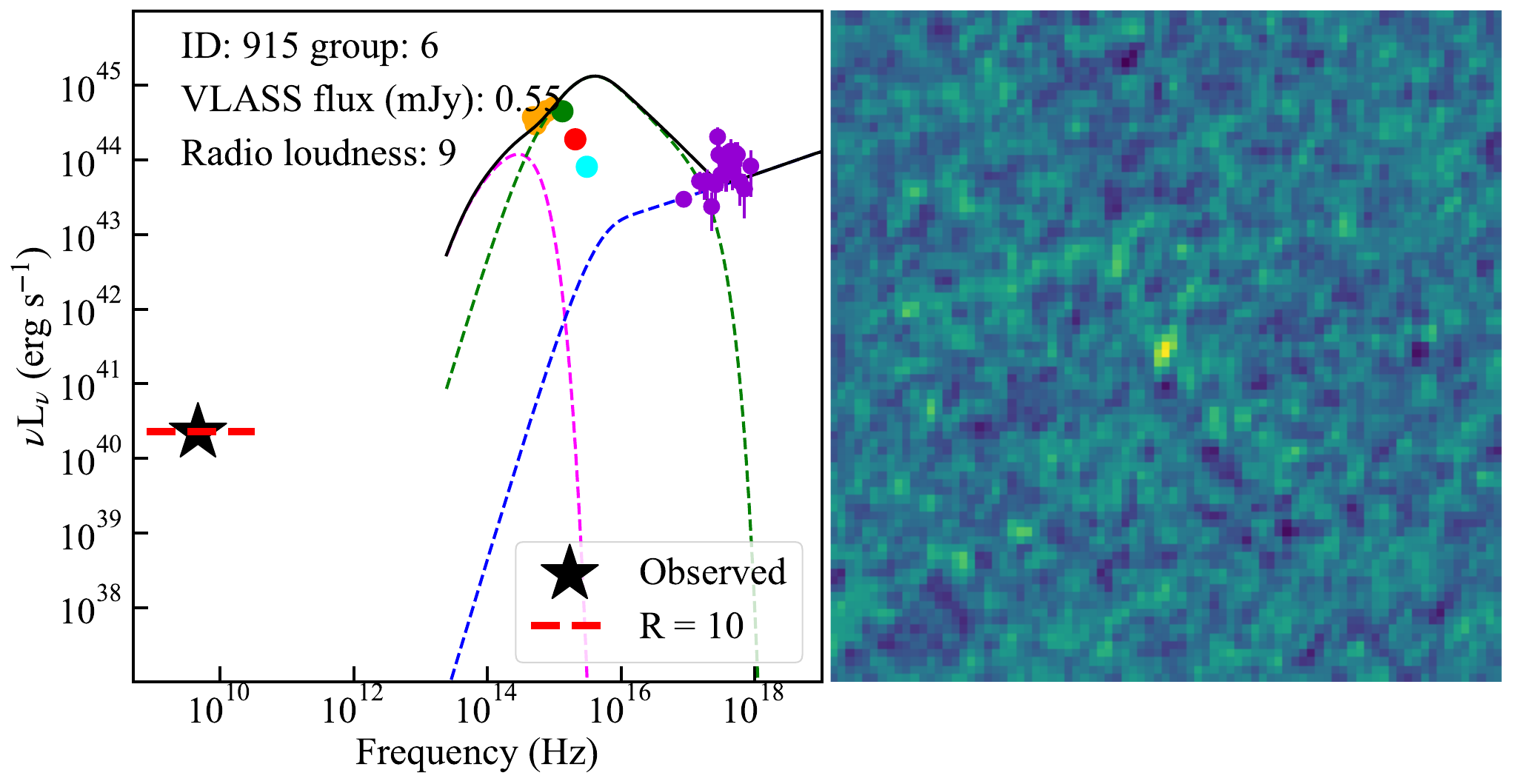}}\\
\subfloat{\includegraphics[width=0.33\textwidth]{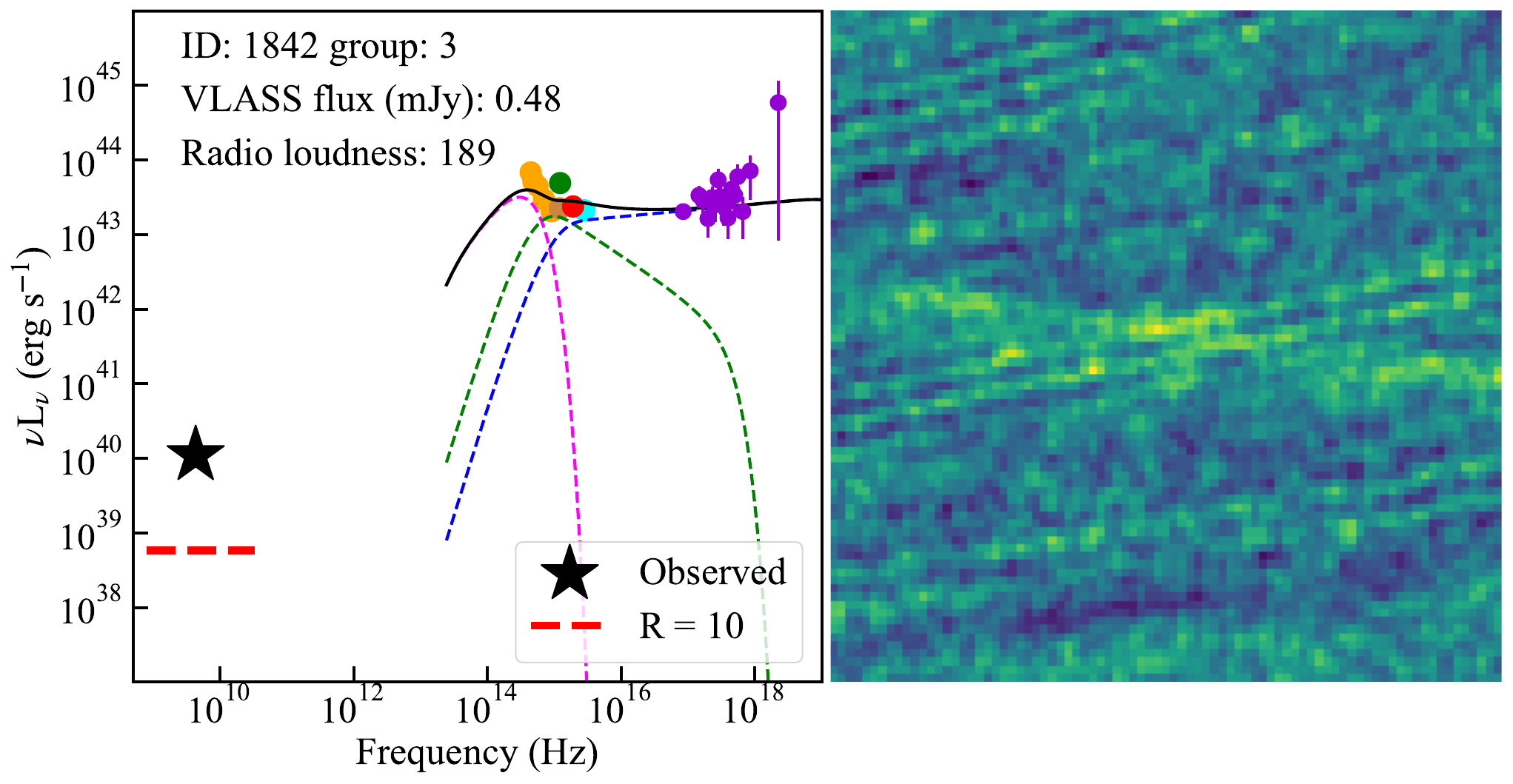}}
\subfloat{\includegraphics[width=0.33\textwidth]{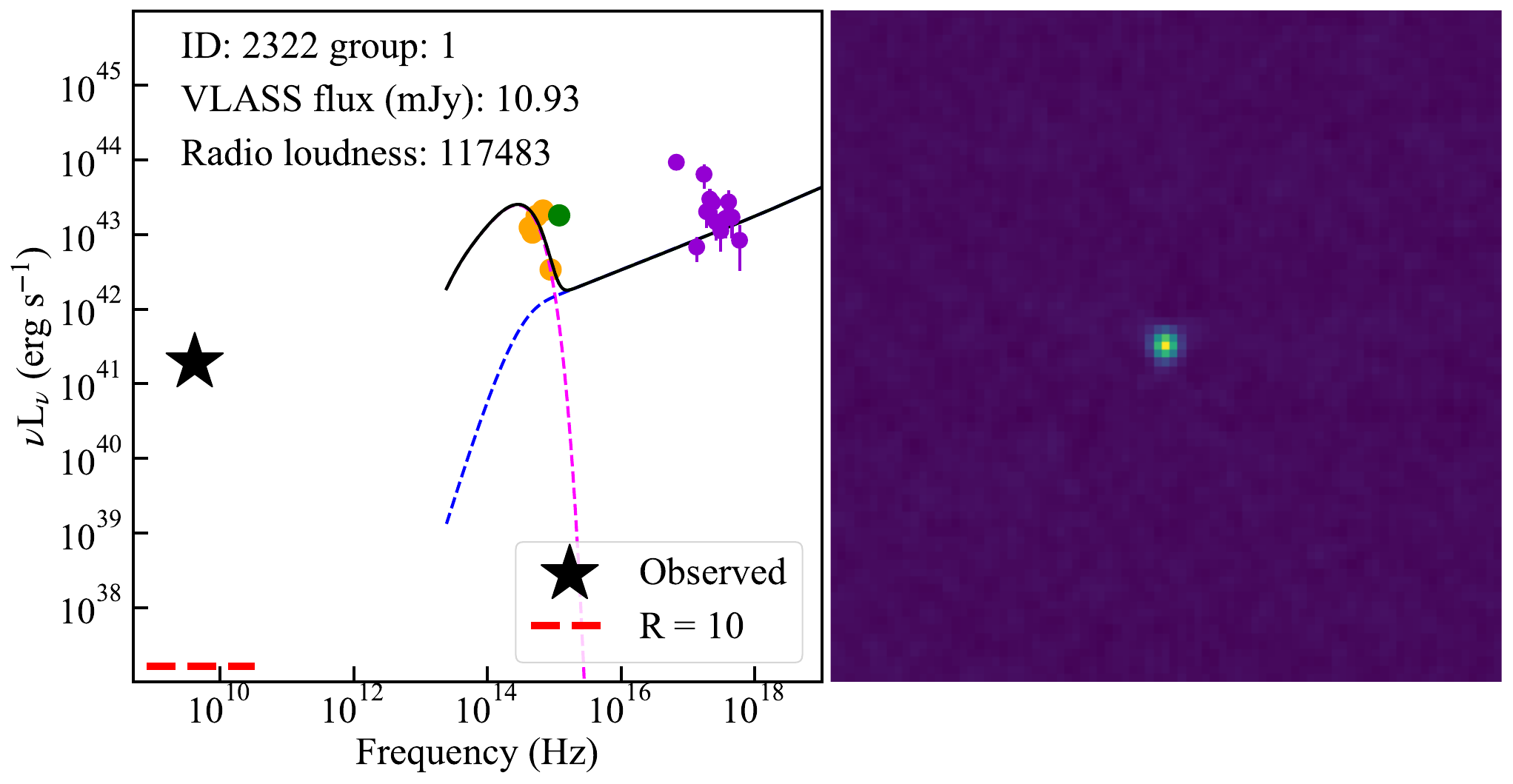}}
\subfloat{\includegraphics[width=0.33\textwidth]{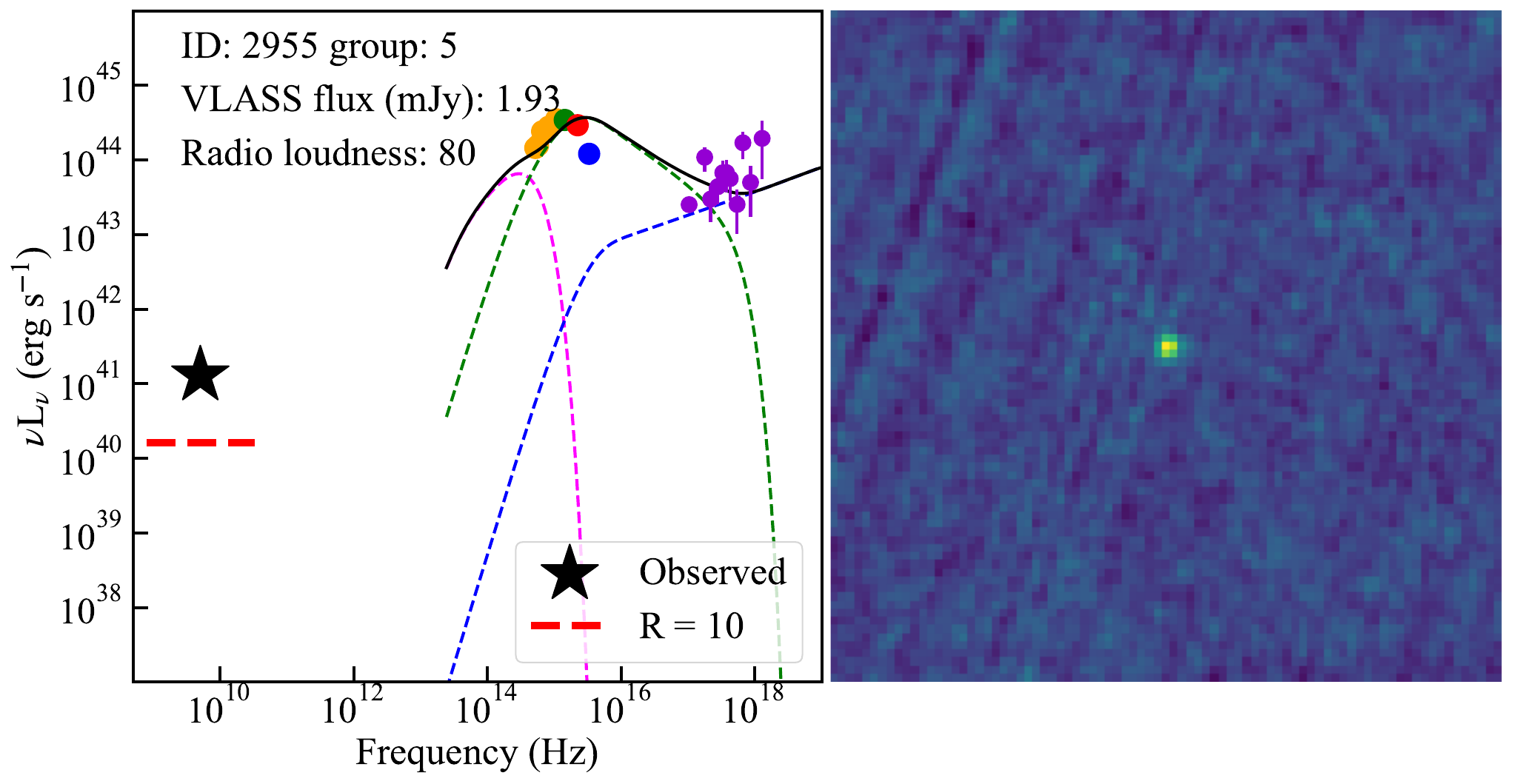}}\\
\subfloat{\includegraphics[width=0.33\textwidth]{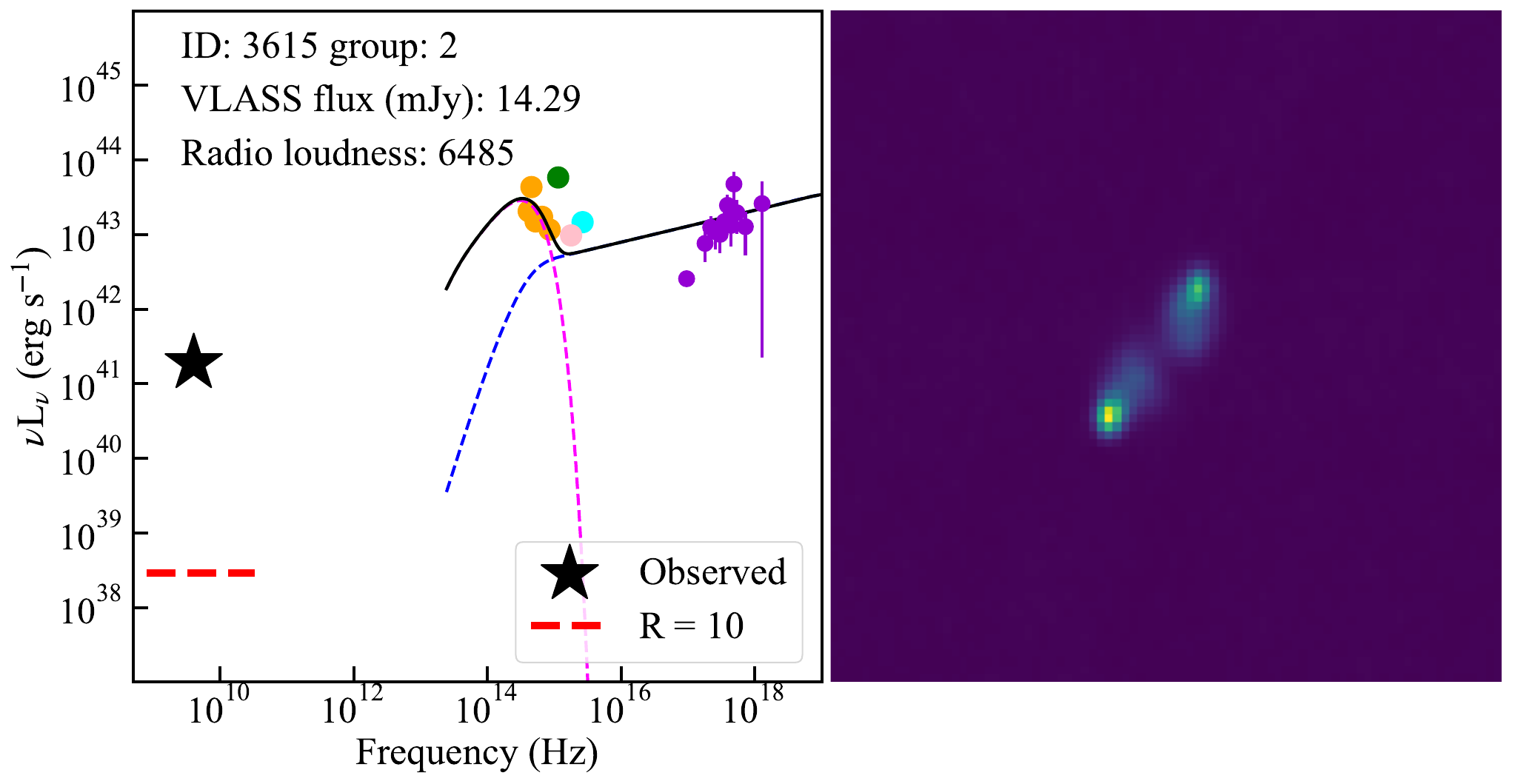}}
\subfloat{\includegraphics[width=0.33\textwidth]{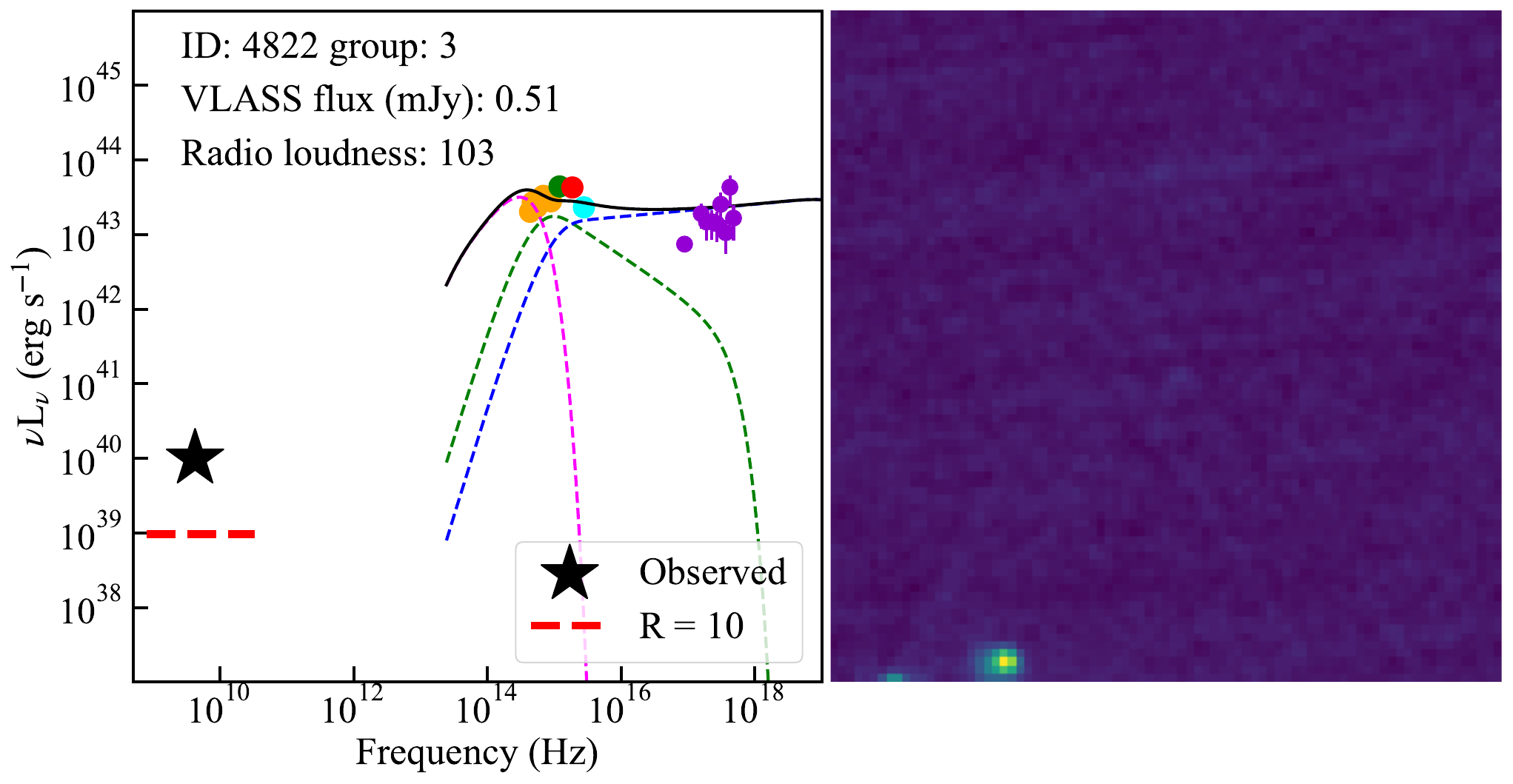}}
\subfloat{\includegraphics[width=0.33\textwidth]{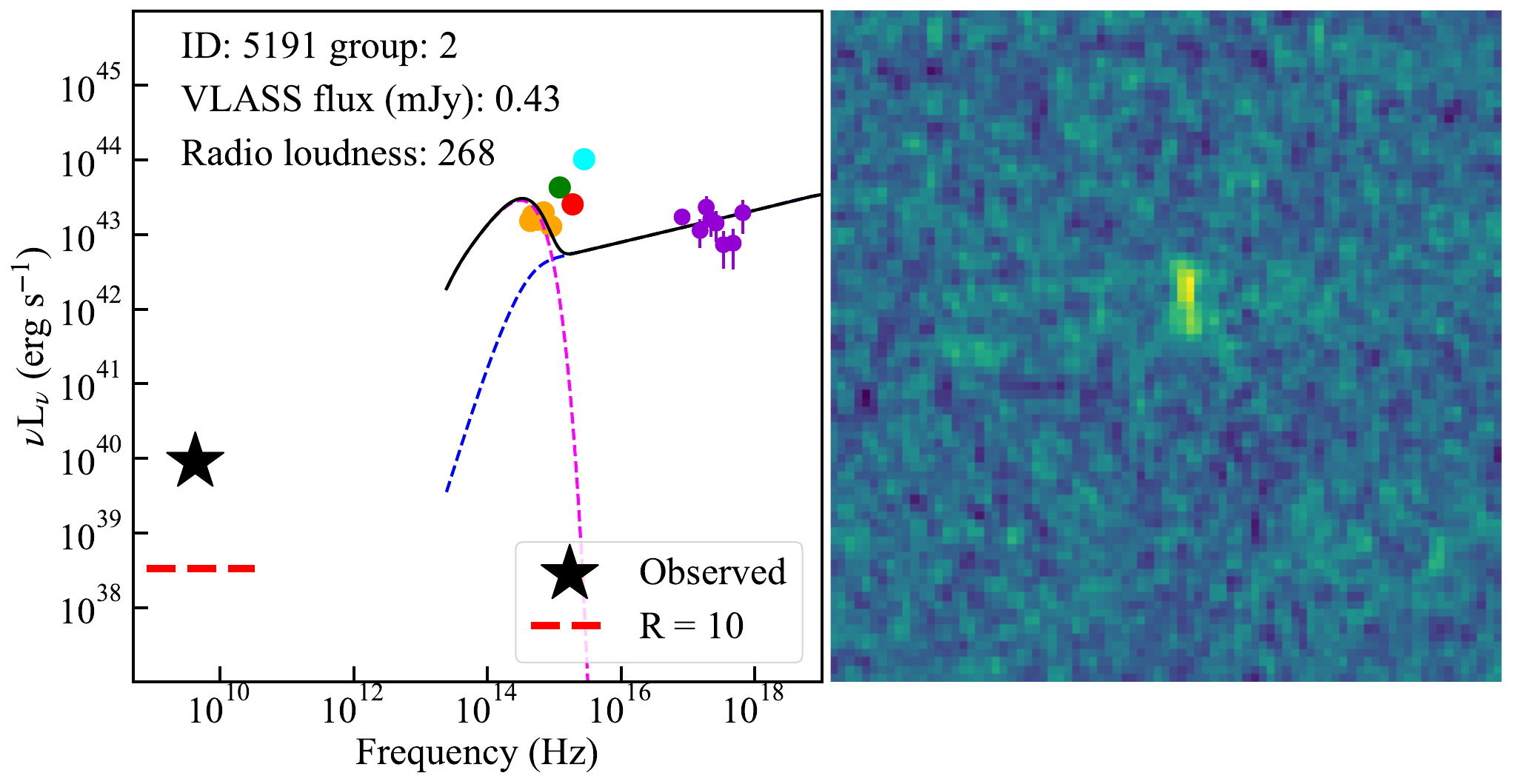}}\\
\subfloat{\includegraphics[width=0.33\textwidth]{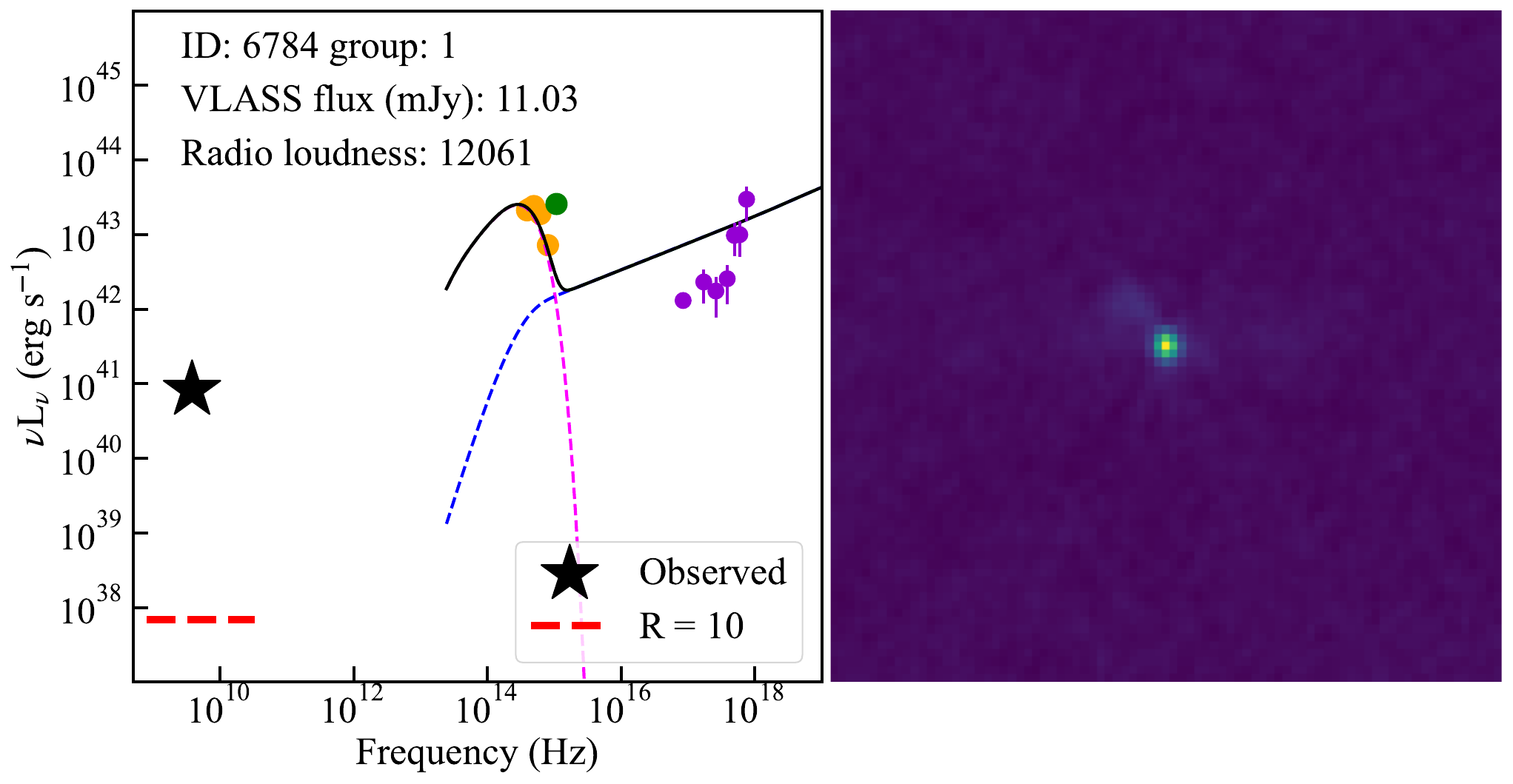}}
\subfloat{\includegraphics[width=0.33\textwidth]{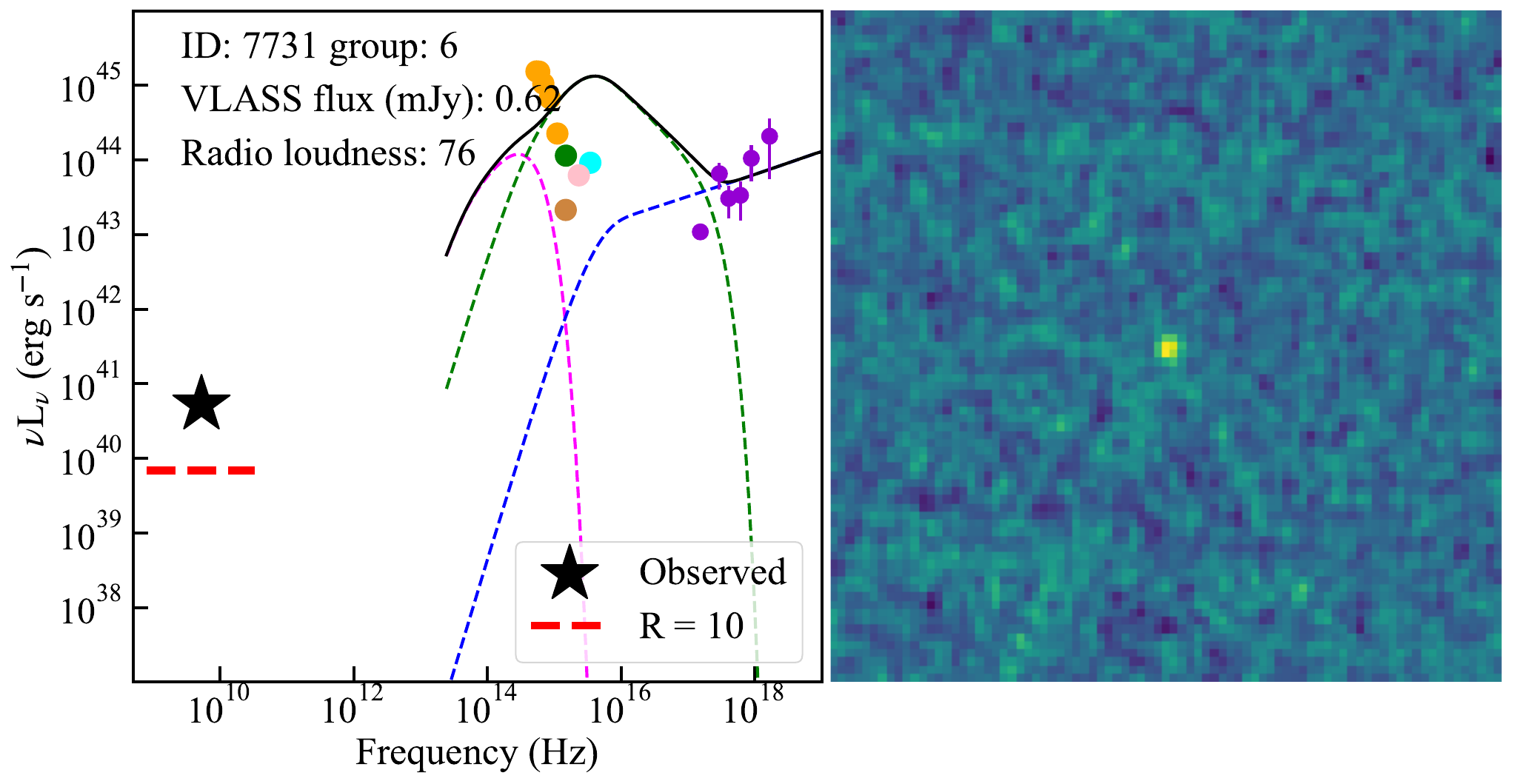}}
\subfloat{\includegraphics[width=0.33\textwidth]{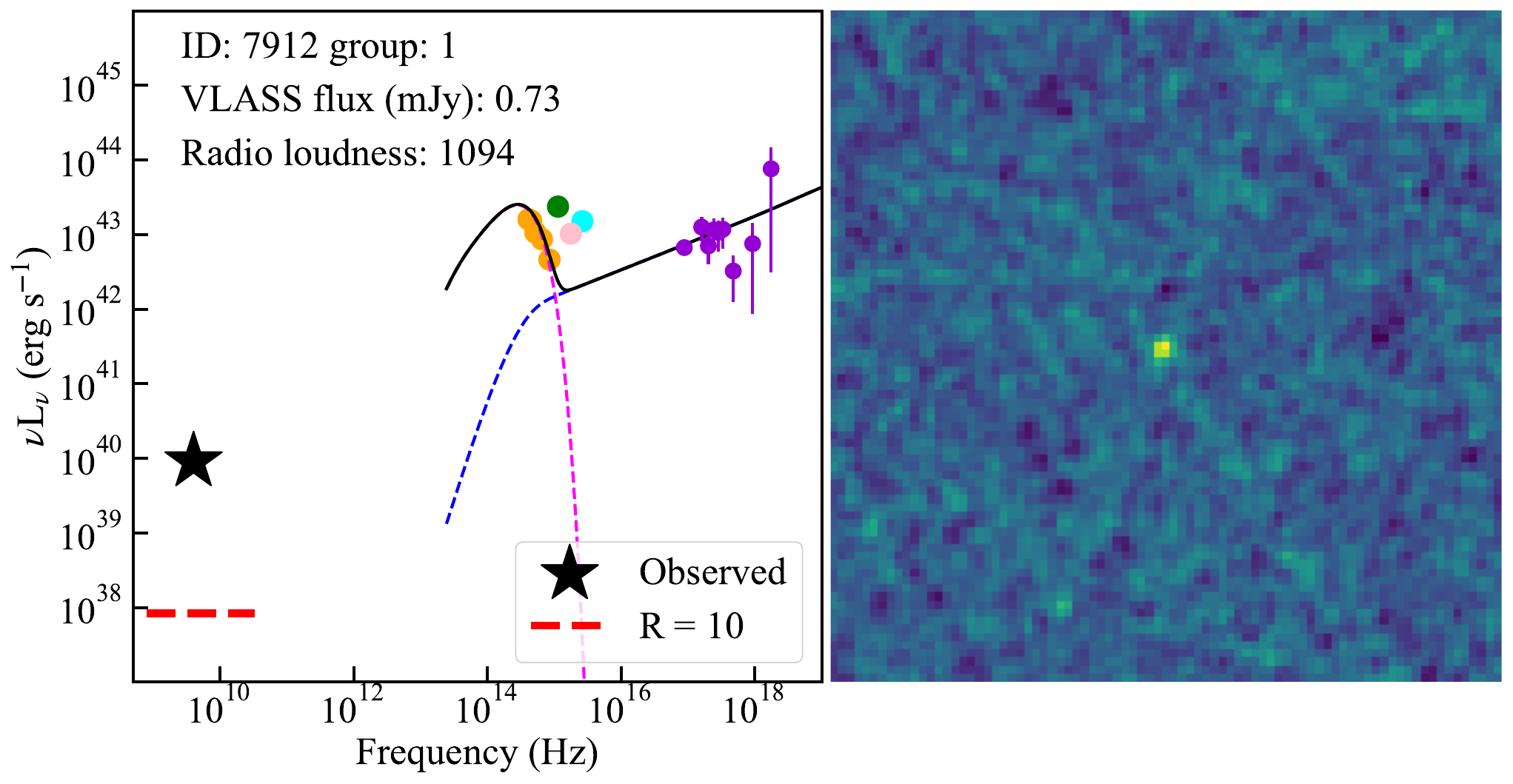}}\\
\subfloat{\includegraphics[width=0.33\textwidth]{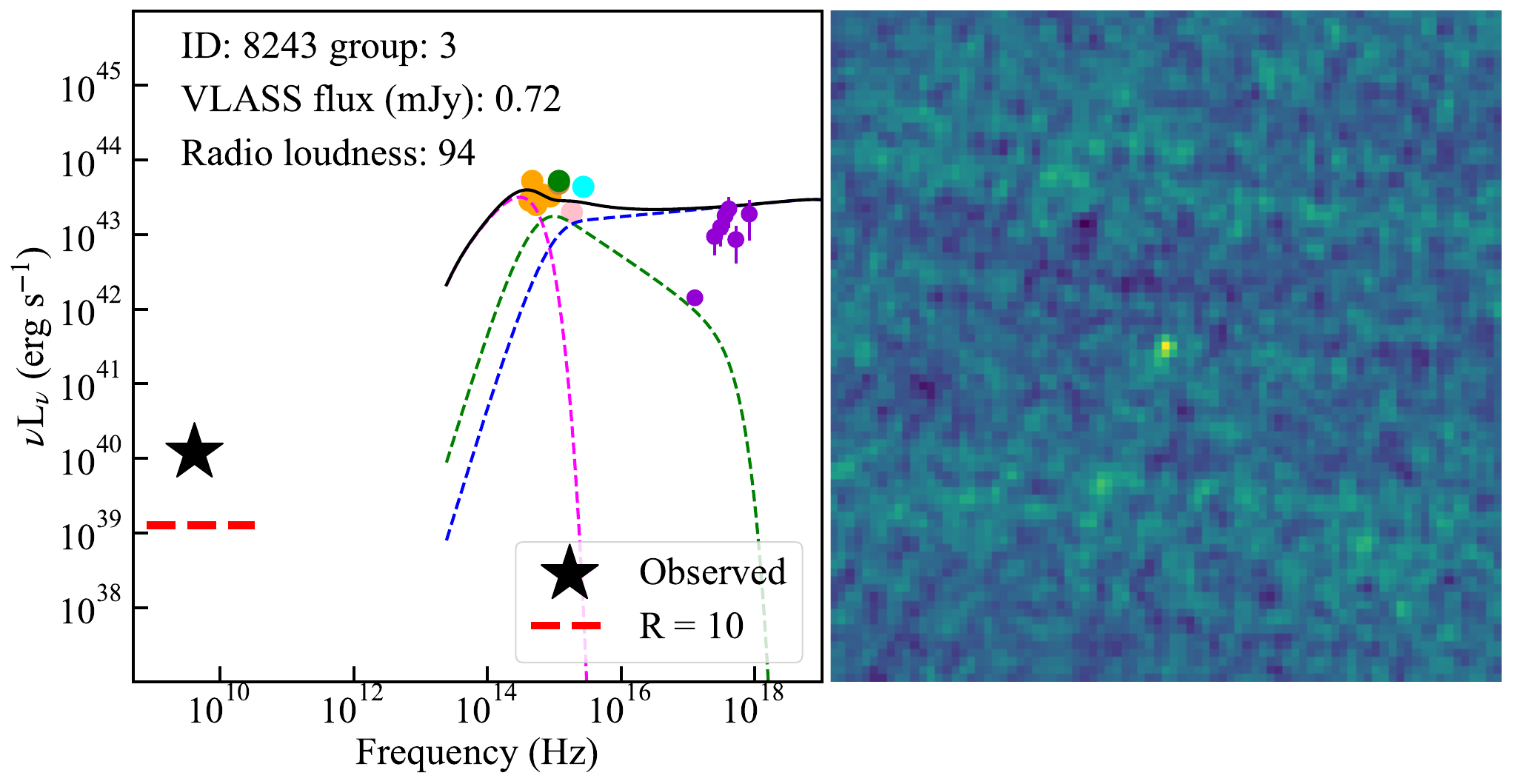}}
\subfloat{\includegraphics[width=0.33\textwidth]{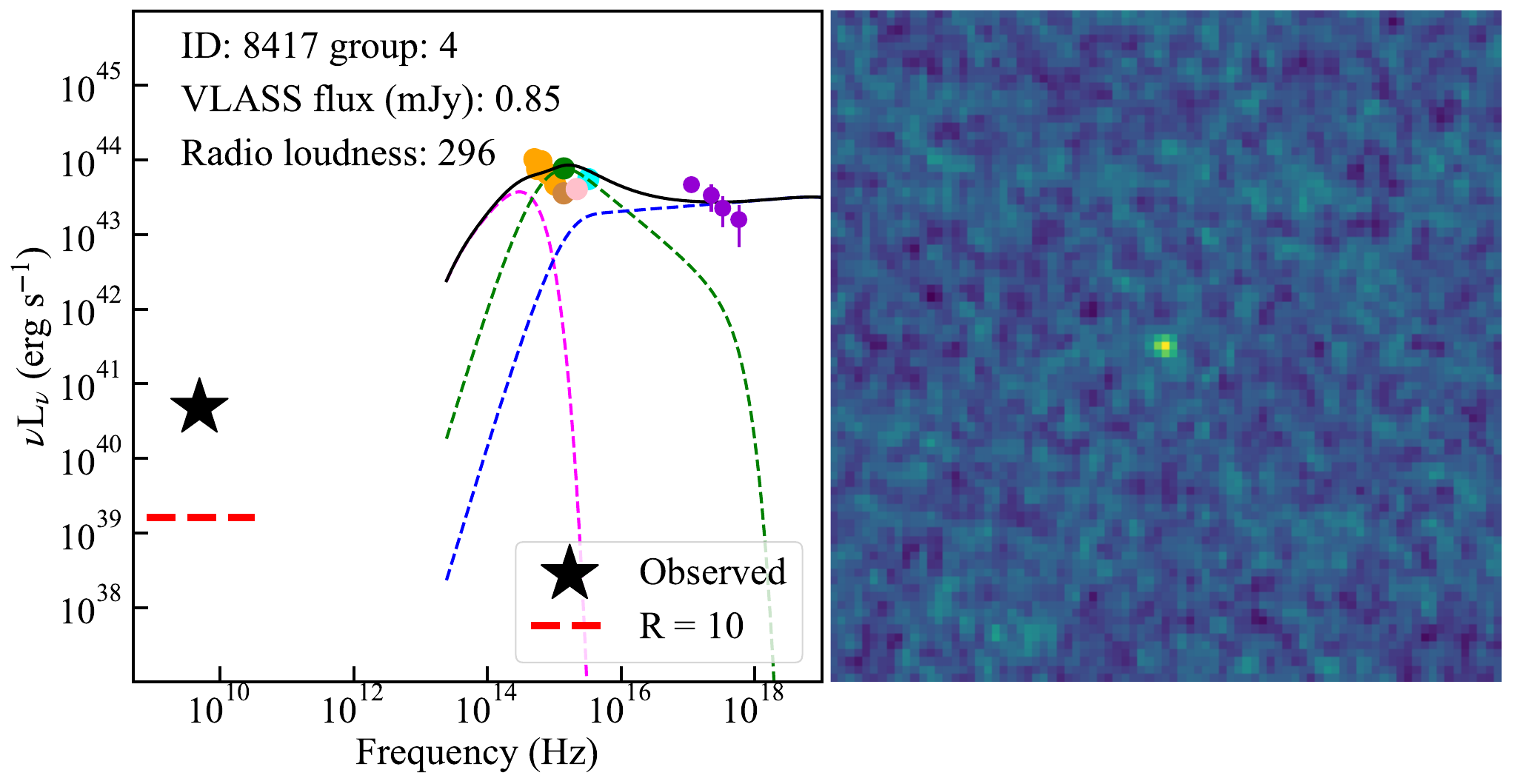}}
\subfloat{\includegraphics[width=0.33\textwidth]{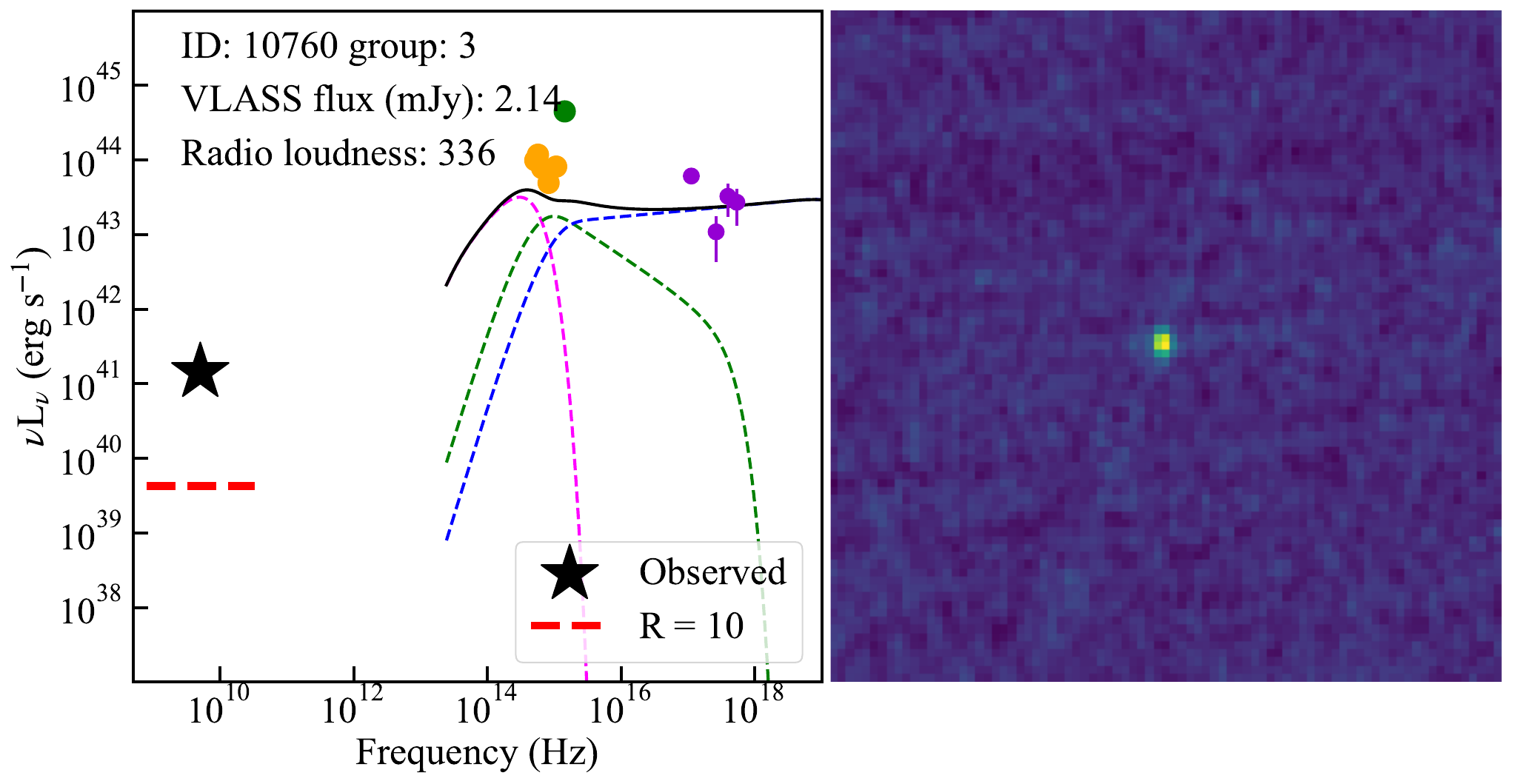}}\\
\subfloat{\includegraphics[width=0.33\textwidth]{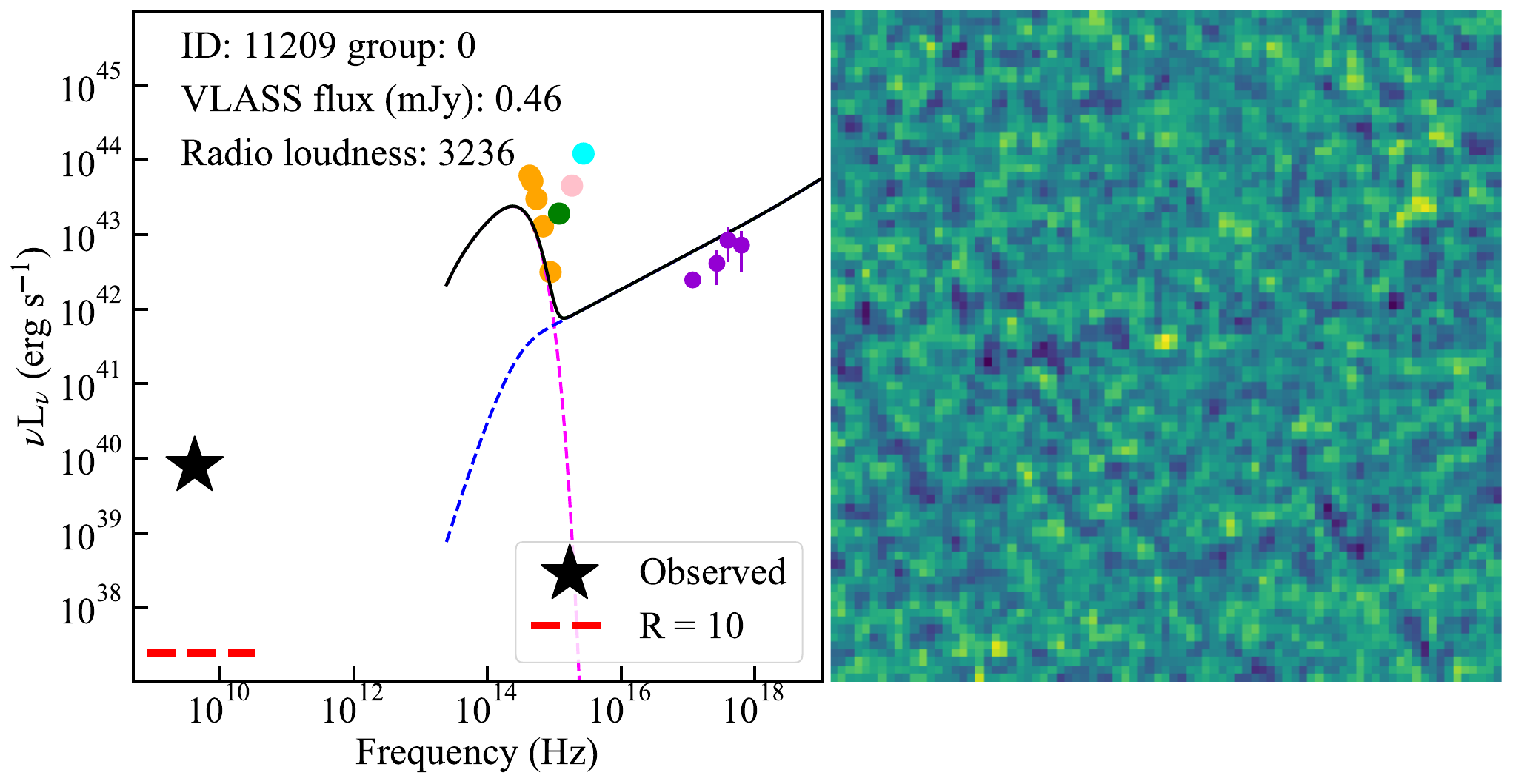}}
\subfloat{\includegraphics[width=0.33\textwidth]{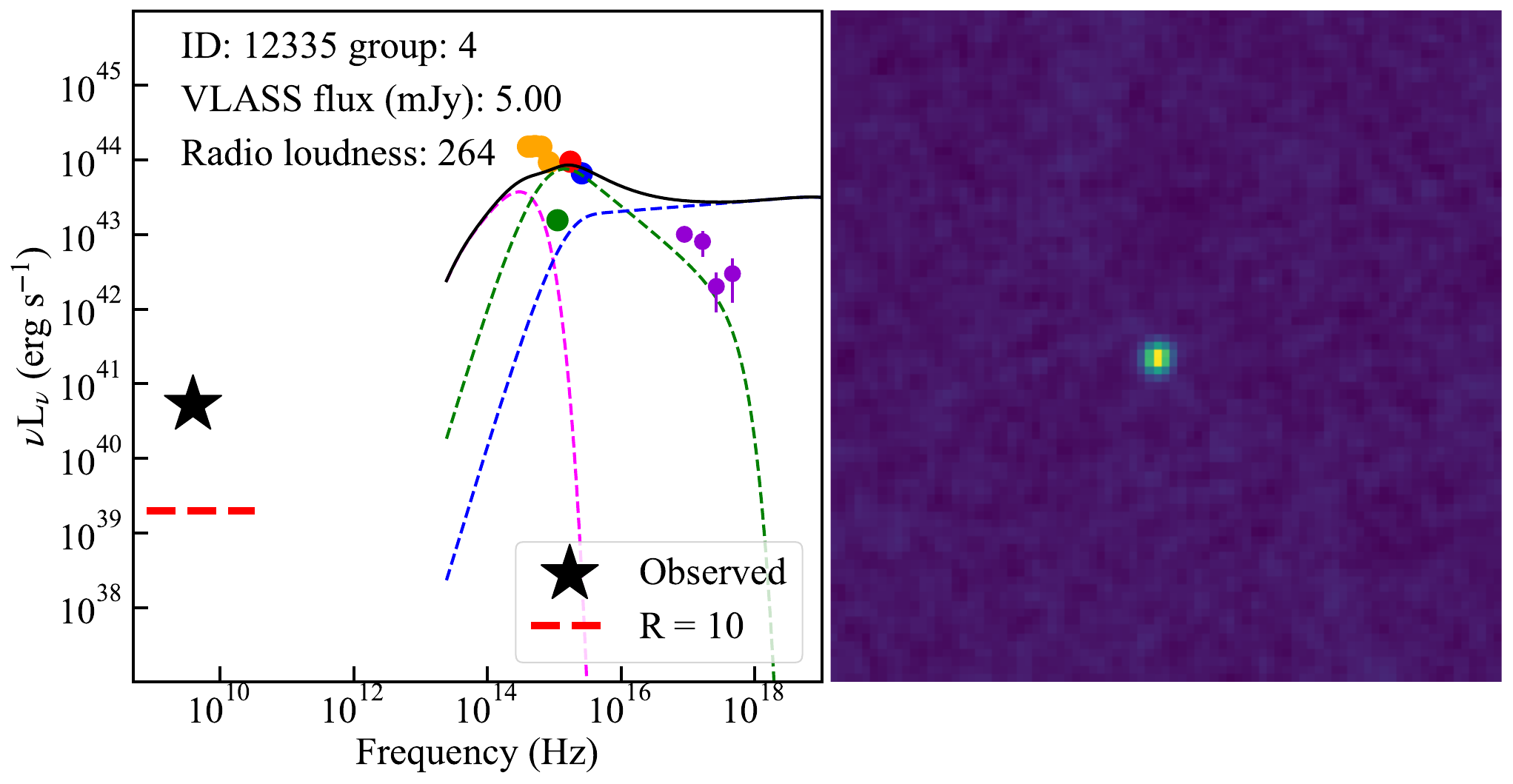}}
\subfloat{\includegraphics[width=0.33\textwidth]{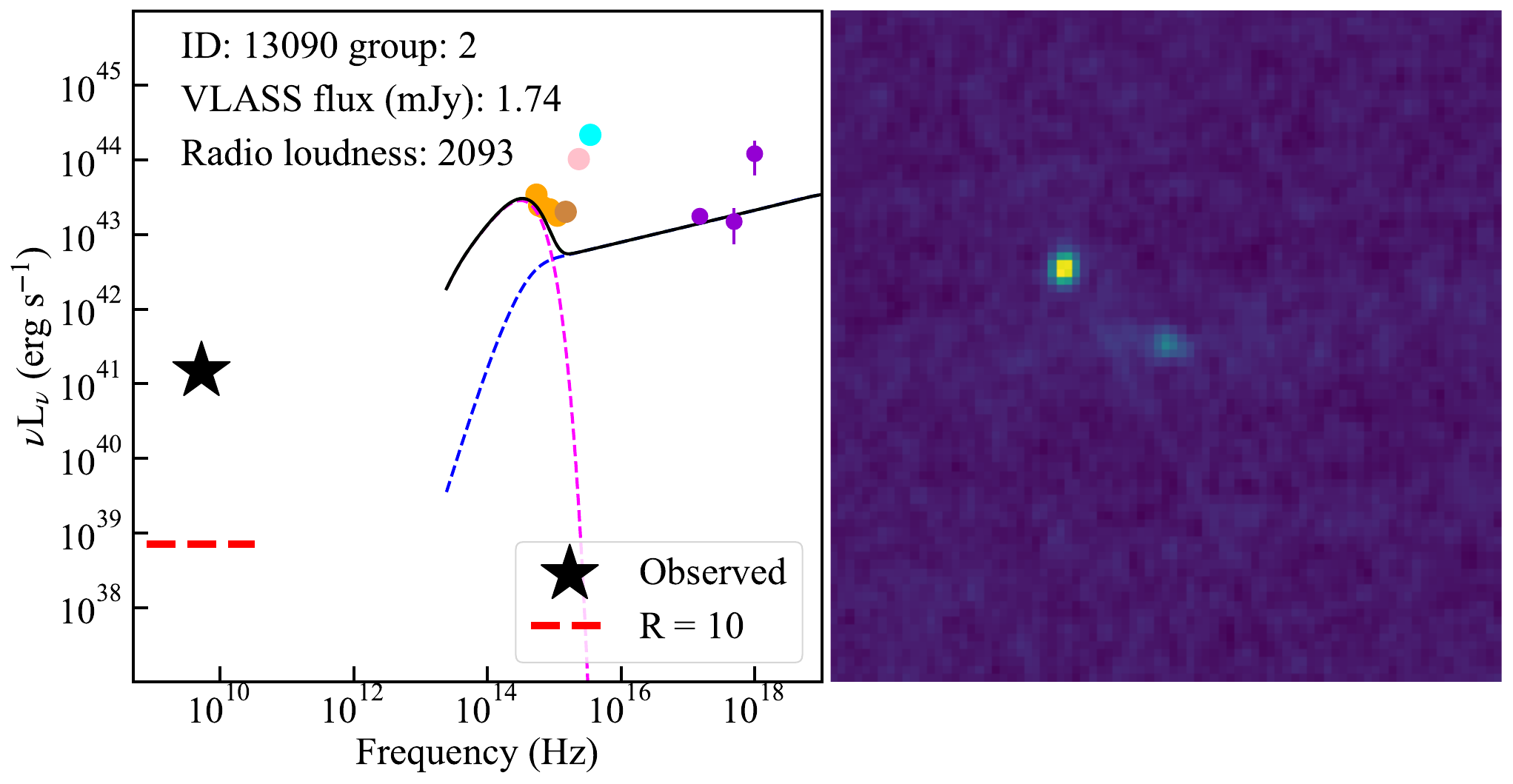}}
\caption{}
\end{figure*}  

\begin{figure*}\ContinuedFloat
\subfloat{\includegraphics[width=0.33\textwidth]{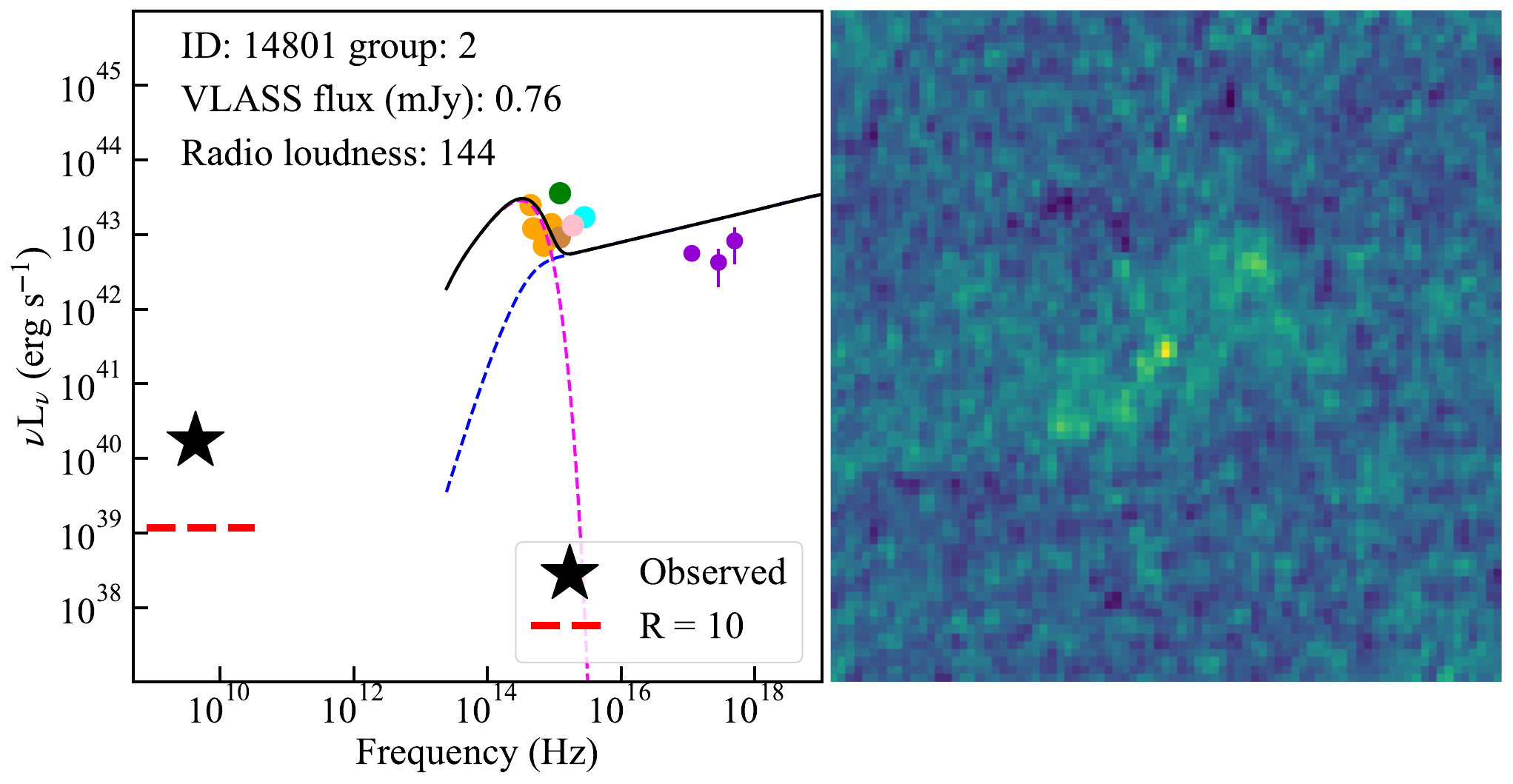}}
\subfloat{\includegraphics[width=0.33\textwidth]{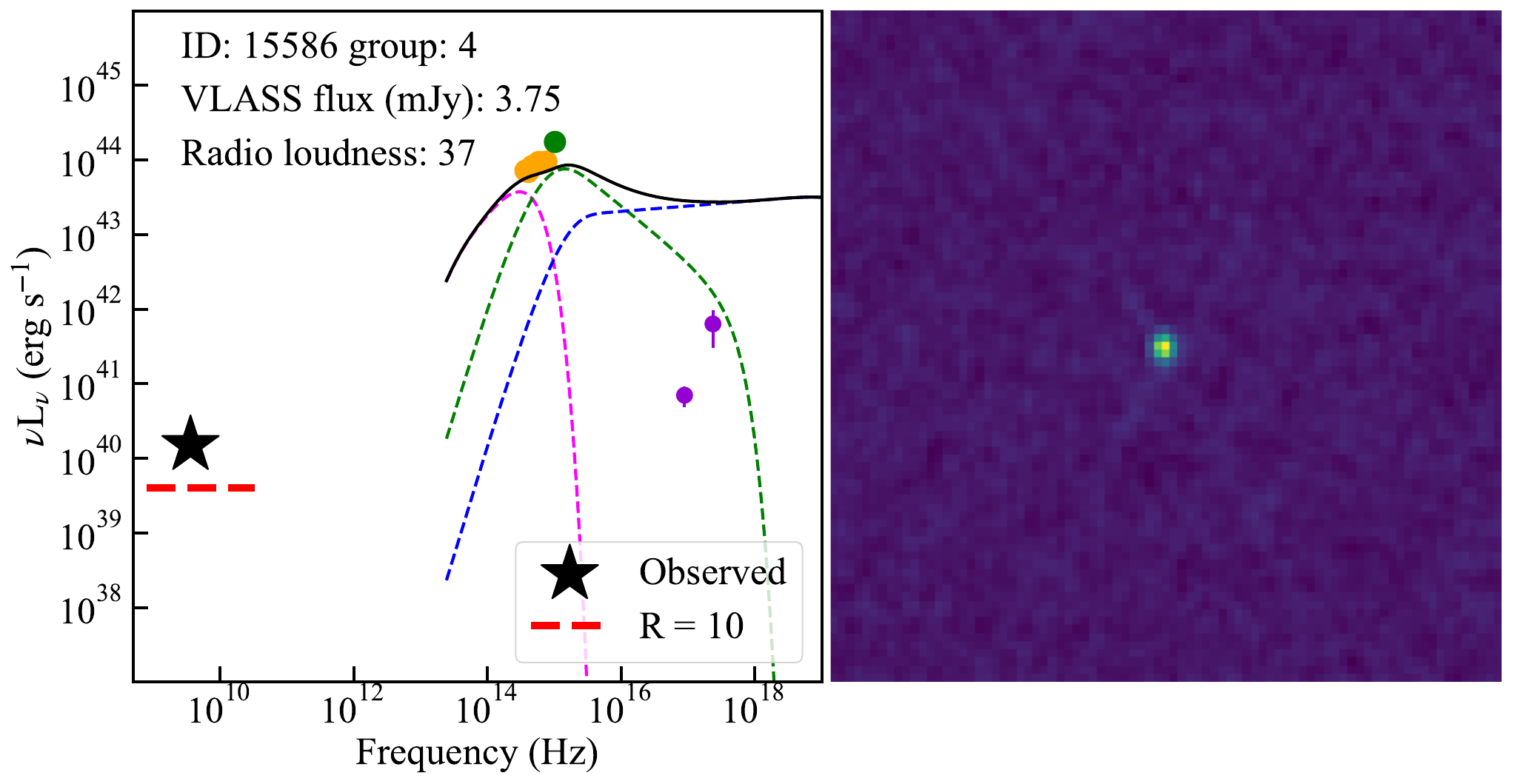}}
\subfloat{\includegraphics[width=0.33\textwidth]{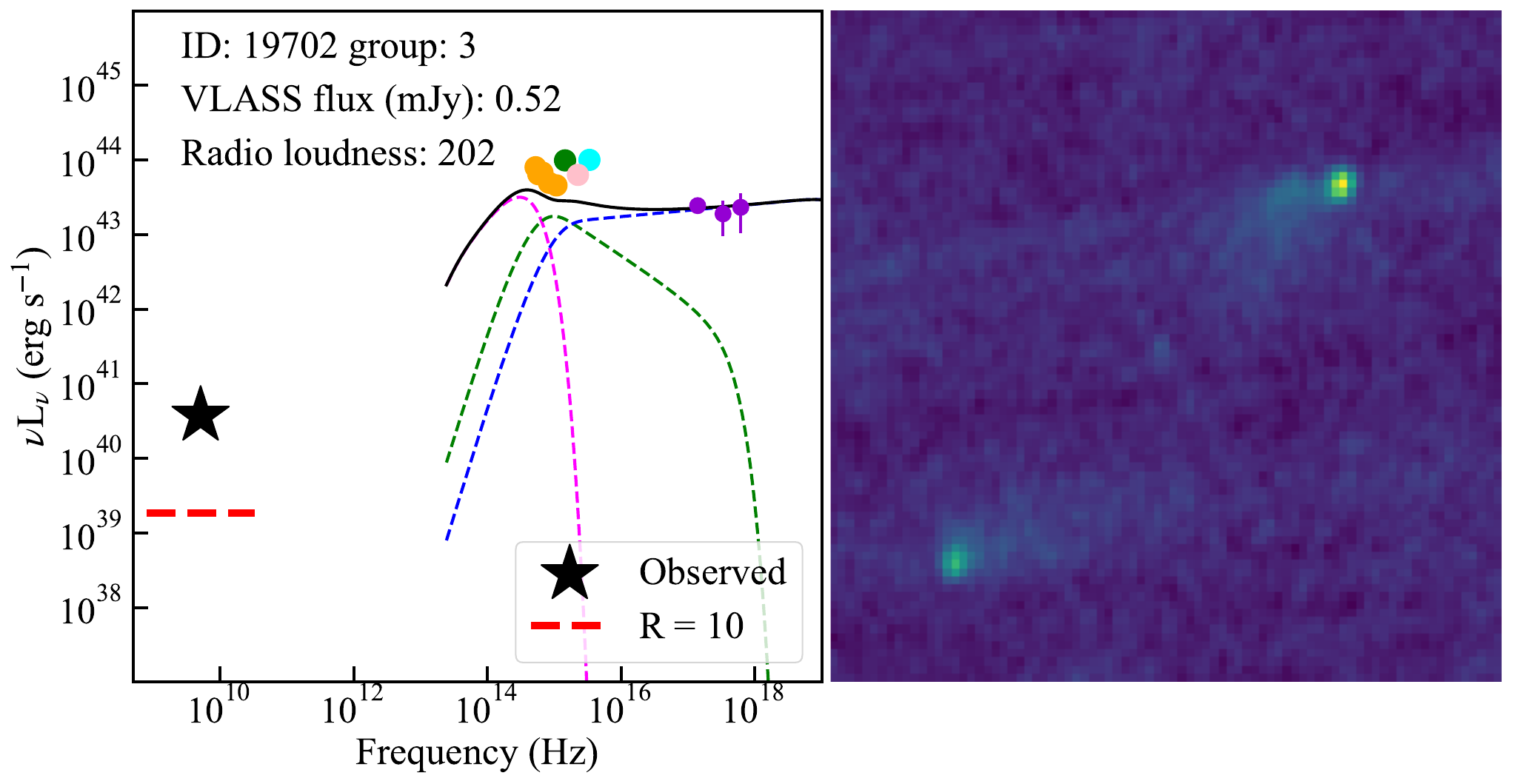}}\\
\subfloat{\includegraphics[width=0.33\textwidth]{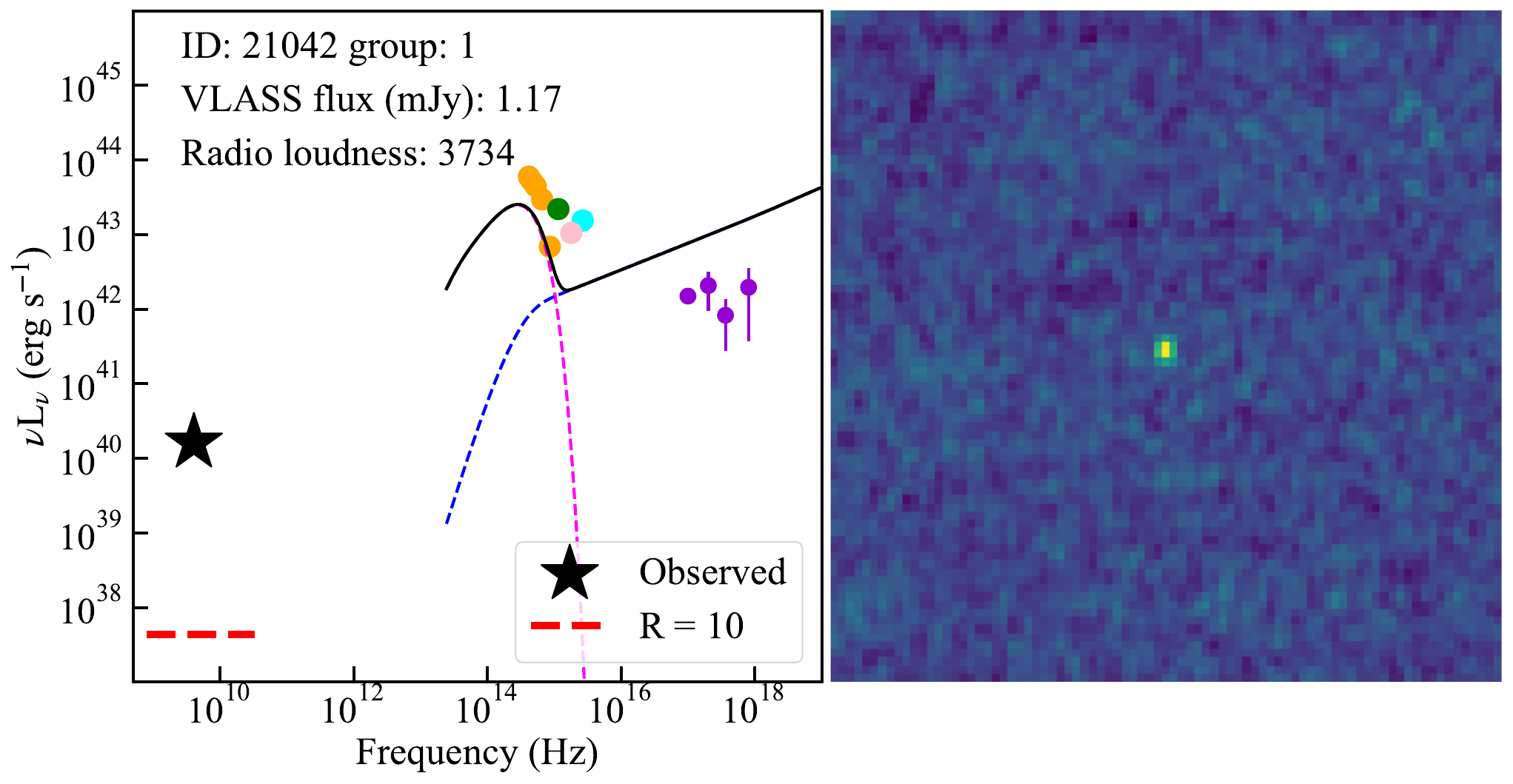}}
\subfloat{\includegraphics[width=0.33\textwidth]{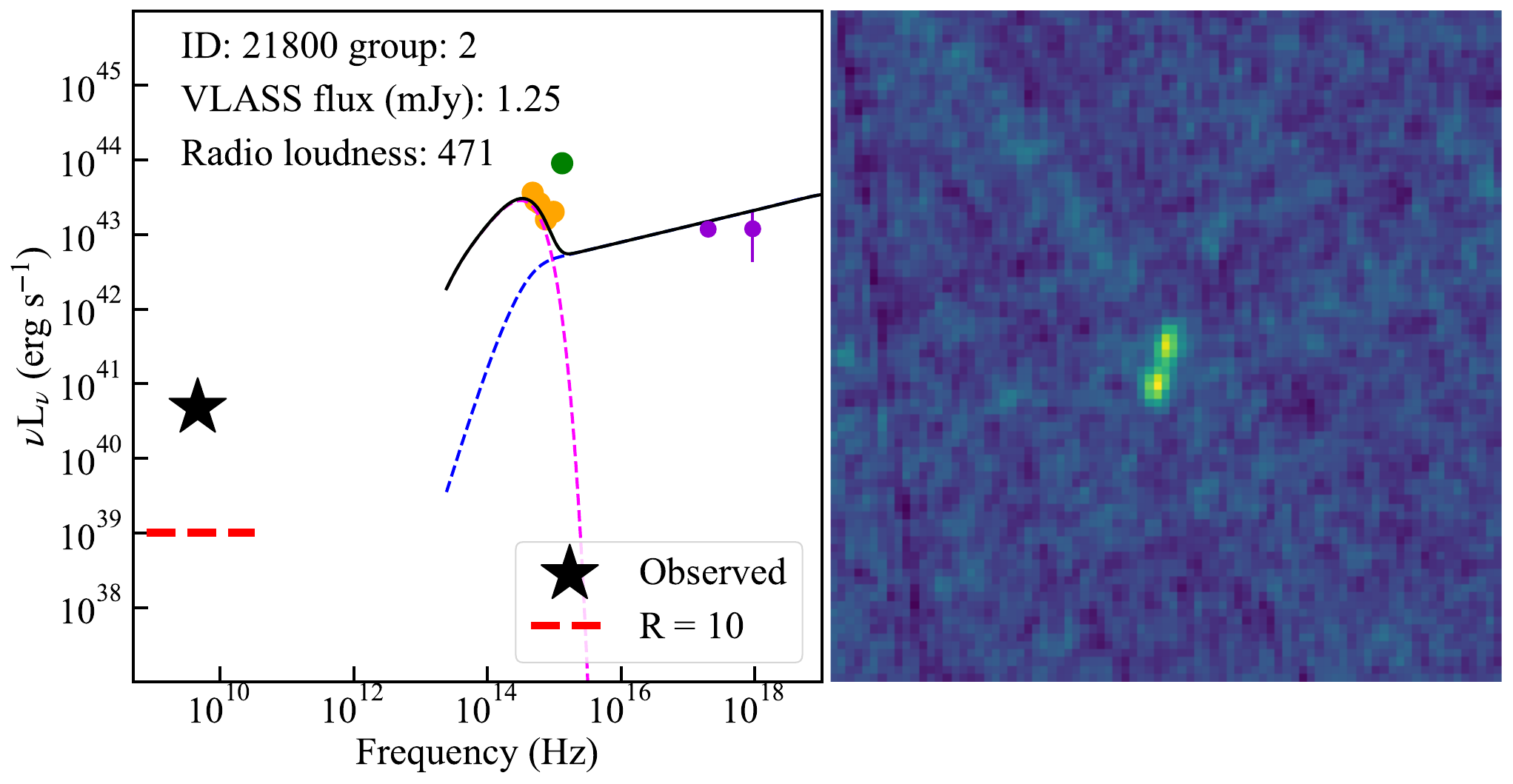}}
\subfloat{\includegraphics[width=0.33\textwidth]{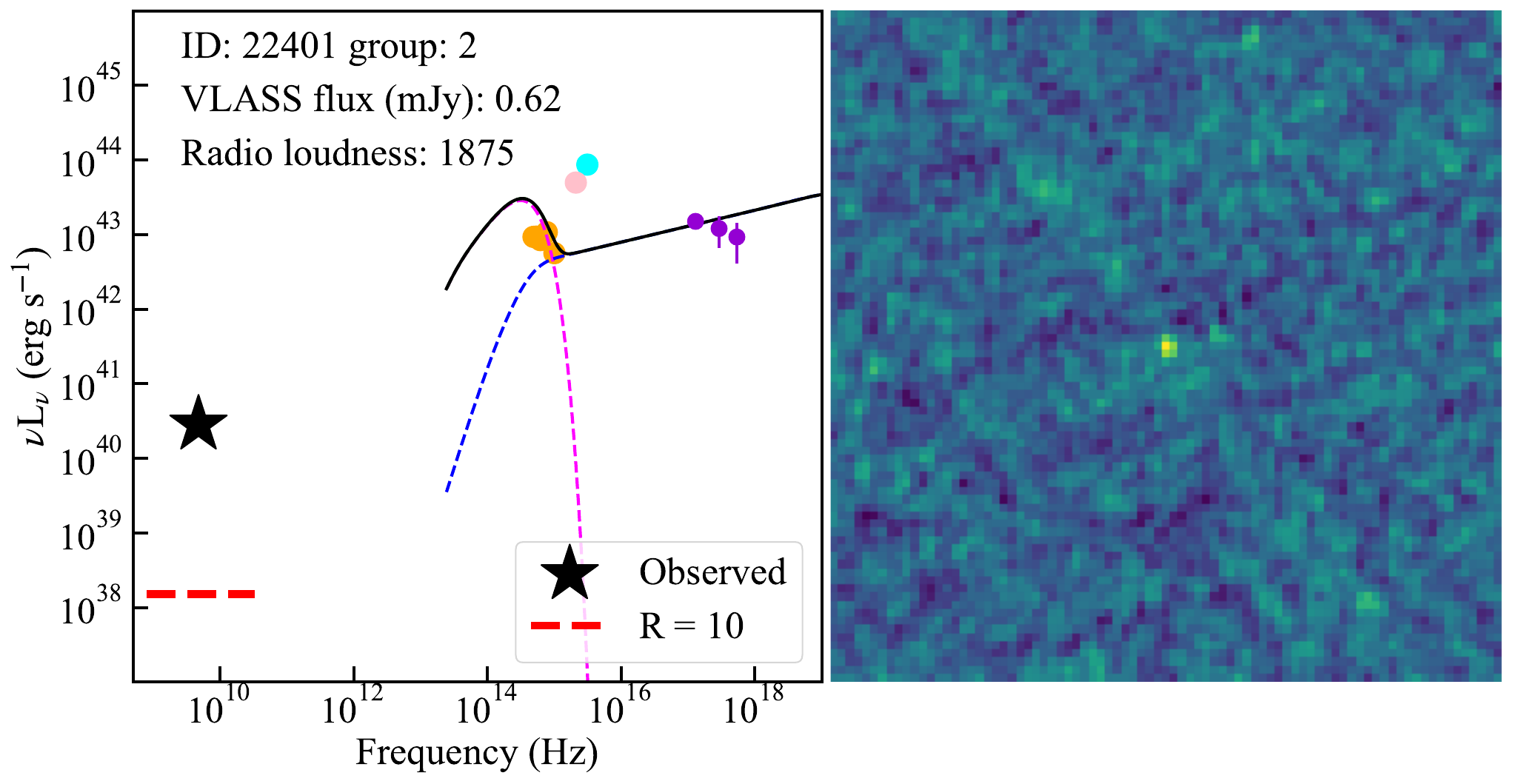}}\\
\subfloat{\includegraphics[width=0.33\textwidth]{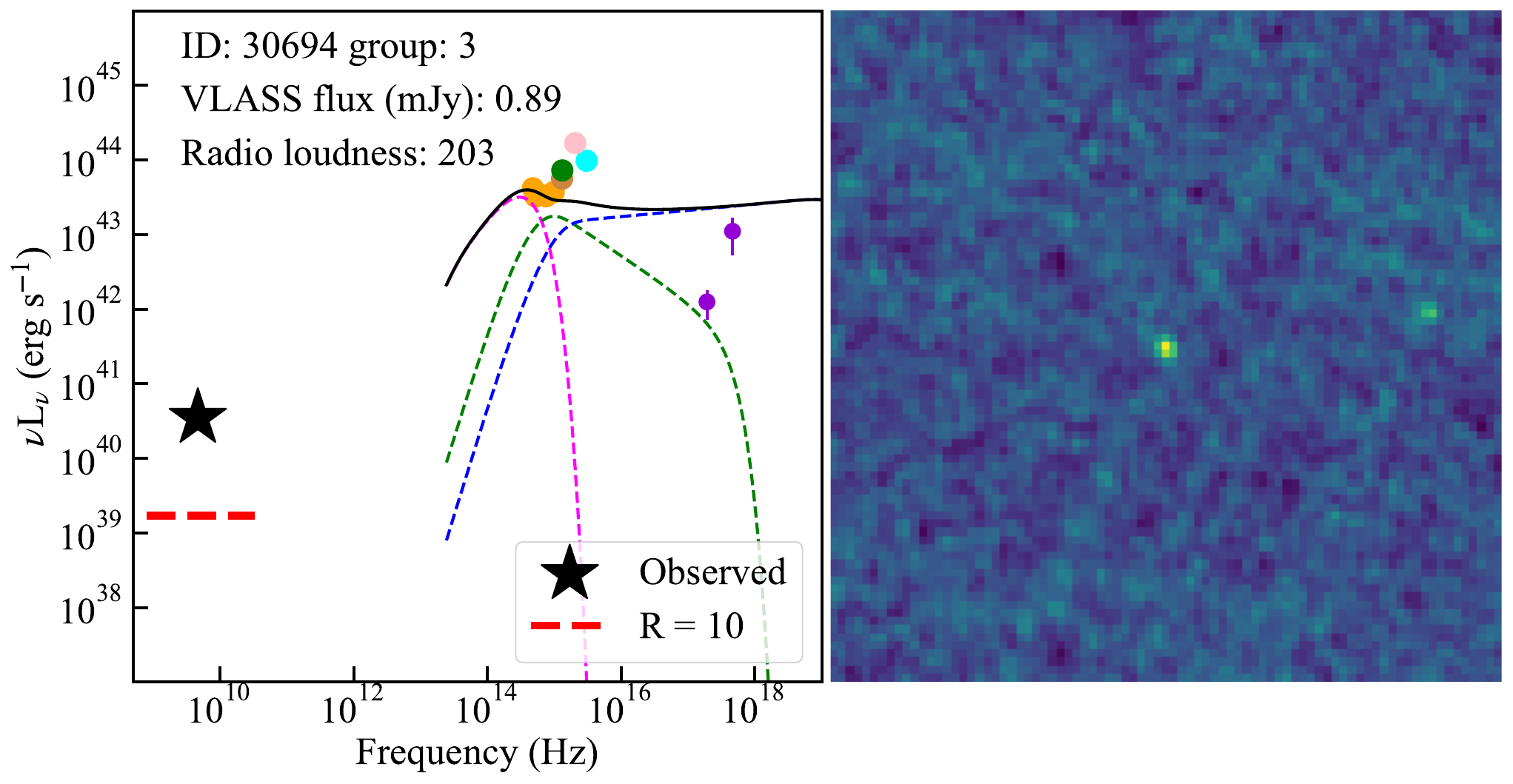}}
\subfloat{\includegraphics[width=0.33\textwidth]{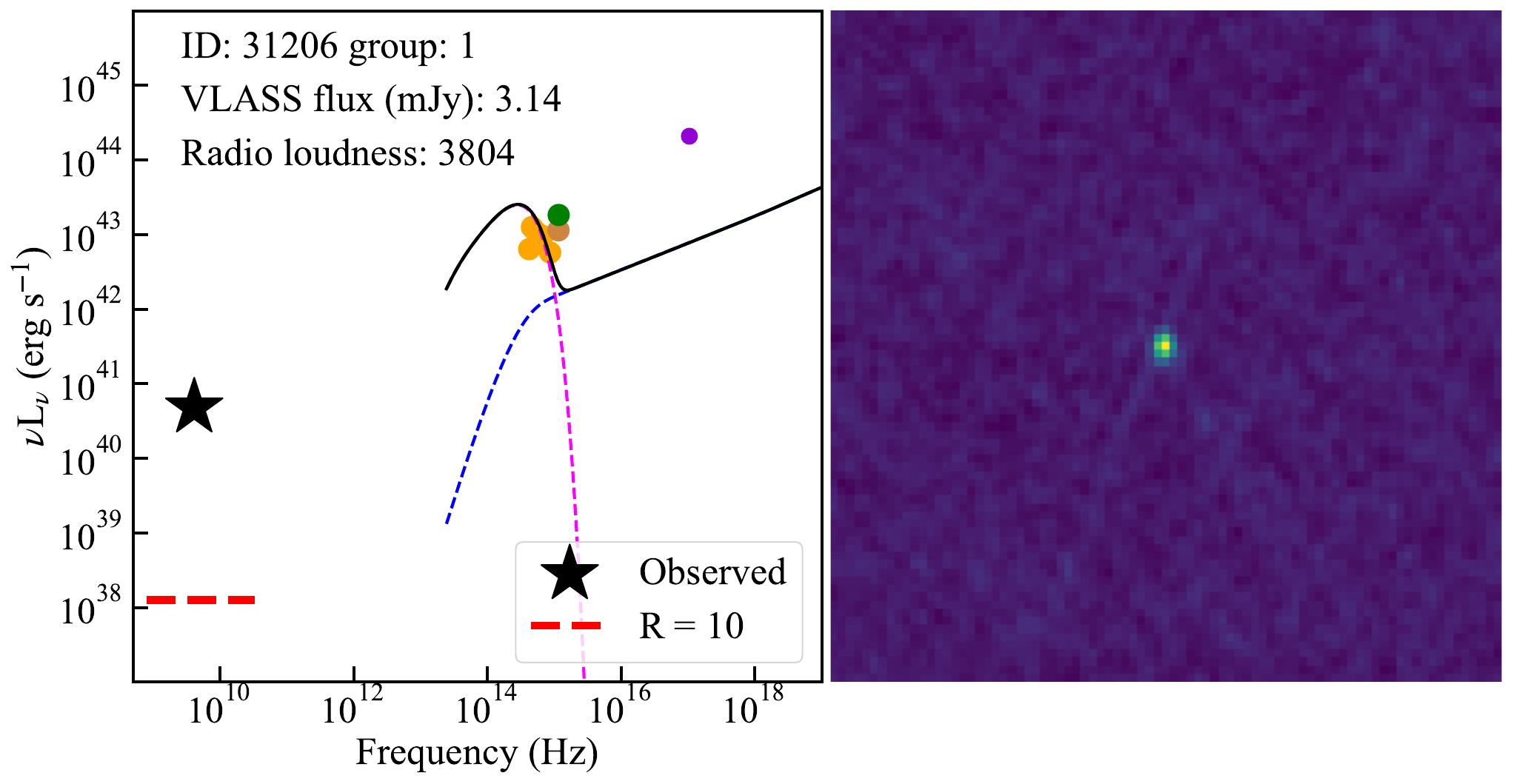}}
\caption{\label{fig:indi_SED} Full SEDs of the individually detected VLASS sources (left) and the VLASS images (right). Pale red (blue) points mark the detected NUV (FUV) photometries of GALEX, while pink (cyan) mark the upper limits. The IDs are the \textit{ID\_SRC} in the eFEDS AGN catalog \citep{Liu_2022}. The group number refers to the bins in \citet{Hagen24}; [0, 1, 2] correspond to the faint, [3, 4] to the middle, and [5, 6, 7] to the bright bins in this paper. Most (21/29) of our sources are compact. There are 5 images showing clear evidence for extended jet lobes (IDs 778 3615 14801 19702 21800) and 4 where there the radio morphology is extended but unclear (IDs 735 1842 5191 6784).  
The radio flux is taken from the single pixel peak central flux, so is weighted to the core flux even where there are obvious extended structures, though the core flux of IDs 3615 is likely contaminated by the extended emission.
The red dashed line on the SED marks $R=10$, the classical definition of a RL source, while the black star shows the level of detected radio luminosity. The VLASS flux limit of $\sim 420\mu$Jy corresponds to $\sim 10^{40}$~ergs~s$^{-1}$ for the mean source redshift here. All the VLASS images have a scale of $81\arcsec\times 81\arcsec$ with pixel size = 1\arcsec.
}
\end{figure*}

\begin{figure*}
\centering
\subfloat{\includegraphics[width=0.8\textwidth]{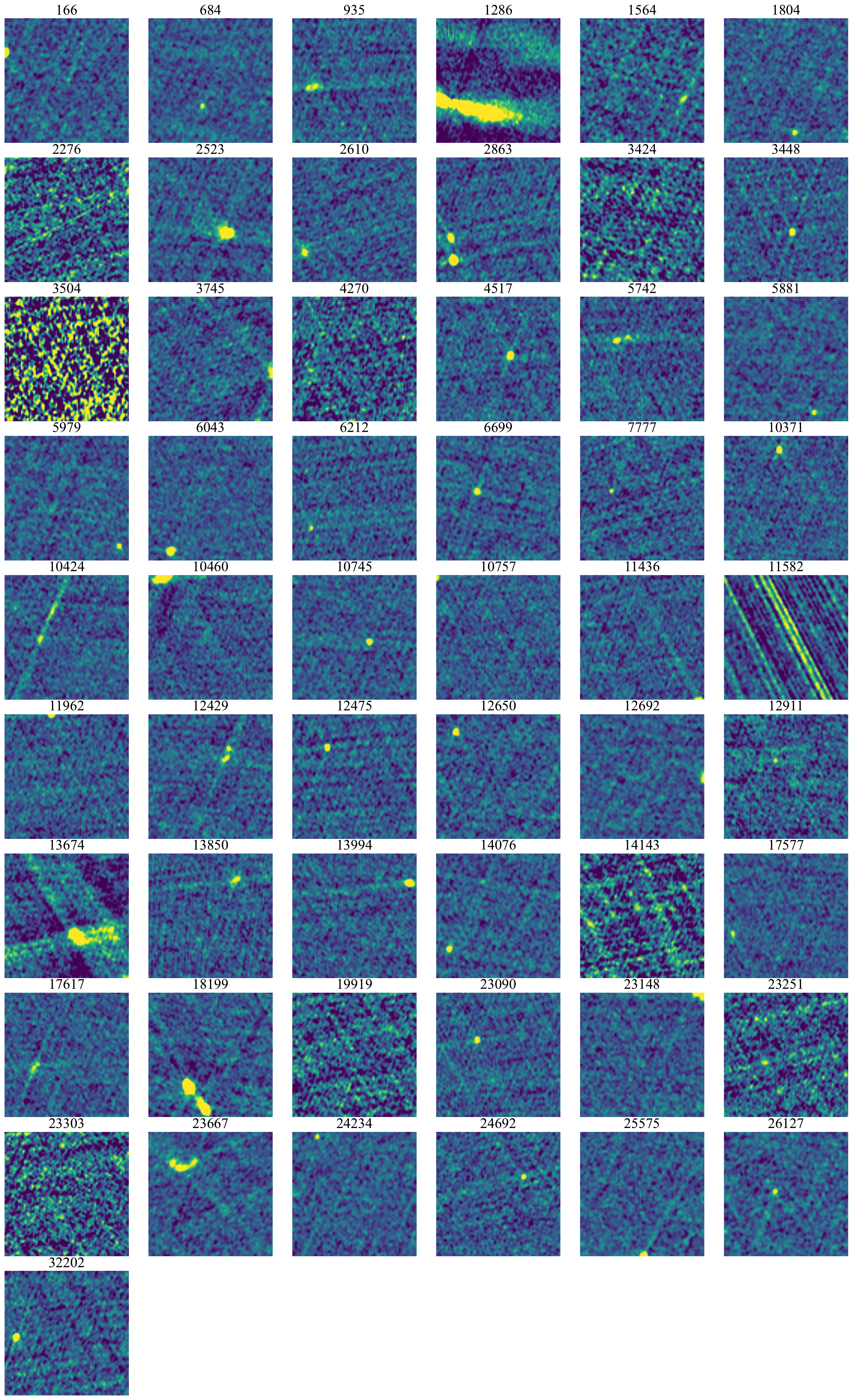}}\\
\caption{\label{fig:vlass_offset} VLASS images of the sources dropped from the sample due to identified counterparts being clearly offset sources and/or image contamination.
}
\end{figure*}


\bsp	
\label{lastpage}
\end{document}